\documentclass{aa}
\usepackage{natbib}
\usepackage{graphicx}
\usepackage{amssymb}
\usepackage{color}

\newcommand{\bl}[1]{\mbox{\boldmath$ #1 $}}

\usepackage{color}

\begin{document}

\title{The circumstellar disk response to the motion of the host star}

\author{Zs. Reg\'aly\inst{1}
    \and
        E. Vorobyov\inst{2,3}
}

\institute{Konkoly Observatory, Research Centre for Astronomy and Earth Sciences, Hungarian Academy of Sciences, 
	    \\H-1121, Budapest, Konkoly Thege Mikl\'os \'ut 15-17, Hungary \\\email{regaly@konkoly.hu}
        \and
        Department of Astrophysics, University of Vienna, 1180, Vienna, Austria \\
        \and
        Research Institute of Physics, Southern Federal University, Stachki Ave. 194, 344090, Rostov-on-Don, Russia
}

\abstract 
{Grid-based hydrodynamics simulations of circumstellar disks are often performed in the curvilinear coordinate system,  in which the center of the computational domain coincides with the motionless star. However, the center of mass may be shifted from the star due to the presence of any non-axisymmetric mass distribution. As a result, the system exerts a non-zero gravity force on the star, causing the star to move in response, which can in turn affect the evolution of the circumstellar disk.}
{We aim at studying the effects of stellar motion on the evolution of protostellar and protoplanetary disks. In protostellar disks, a non-axisymmetric distribution of matter in the form of spiral arms 
and/or massive clumps can form due to gravitational instability. Protoplanetary disks can also feature non-axisymmetric structures caused by an embedded high-mass planet or a large-scale vortex formed at viscosity transitions.}
{We use 2D grid-based numerical hydrodynamic simulations to explore the effect of stellar motion. We adopt a non-inertial polar coordinate system centered on the star, in which the stellar motion is taken into account by calculating the indirect potential caused by the non-axisymmetric disk, a high-mass planet, or a large-scale vortex. We compare the results of numerical simulations with and without stellar motion.}
{We found that the stellar motion has a moderate effect on the evolution history and the mass accretion rate in protostellar disks, reducing somewhat the disk size and mass, while having a profound effect on the collapsing envelope, changing its inner shape from an initially axisymmetric to a non-axisymmetric configuration. 
Protoplanetary disk simulations show that the stellar motion slightly reduces the width of the gap opened by a high-mass planet, decreases the planet migration rate, and strengthens 
the large-scale vortices formed at the viscosity transition.}
{We conclude that the inclusion of the indirect potential is recommended in grid-based hydrodynamics simulations of circumstellar disks which use the curvilinear coordinate system.}
  

\authorrunning{Zs. Reg\'aly and E. Vorobyov}
\titlerunning{Stellar wobbling}
\maketitle

\section{Introduction}

The formation mechanisms of stars and planets are intrinsically connected. The stars form owing 
to the gravitational collapse of molecular clouds and  circumstellar disks develop 
around the forming stars thanks to the conservation of angular momentum. 
The circumstellar disk, also known as the protostellar disk, 
is embedded in the parental nebula in the early phase, in which the mass accretion from 
the parental nebula onto the disk exceeds the mass loss from the disk to the star.  
If the protostellar disk grows sufficiently large 
(about $\ga0.07\,M_\odot$) it becomes gravitationally unstable, often resulting in the 
formation of gravitationally bound clumps. As the parental nebula is consumed, the nebular 
accretion ceases and the protostellar disk evolves into the protoplanetary disk, which is 
thought to be the birth place for planets. 

The evolution of gaseous protostellar and protoplanetary disks and their interactions with the 
newly born planets can be studied by means of hydrodynamical models. In hydrodynamical simulations, one solves the fluid dynamics equations describing the time evolution of mass density, momentum, 
and energy. Since the advection term in the momentum equation makes the system of equations non-linear, the analytical investigations are applicable only for highly idealized flows. Therefore, numerical methods have been developed  since the mid-1960s to solve the system of hydrodynamical equations numerically.

Nowadays, there are two widely used different approaches to solving those differential equations numerically:  smoothed-particle hydrodynamics (SPH) and grid-based hydrodynamics. In the SPH method 
(see a review of \citealp{Springel2010}) the fluid is handled as a set of discrete elements, referred to as particles, while in the grid-based approach  (see a review of \citealp{Teyssier2015}) the flow is handled with discrete values of fluid properties on a numerical grid. 

In astrophysical models that are designed to study the formation of stars and planets, gravitational interaction naturally enters the system of hydrodynamical equations. If 
any non-axisymmetric mass distributions are present (such as massive clumps and spirals 
in gravitationally unstable protostellar disks or massive planets and large-scale vortices 
in protoplanetary disks), the center of mass of the star plus disk 
system (or its barycenter) does not coincide with the position of the star.  
This leads to the motion of the star
in response to the net gravitational force of the disk and planets.

In the SPH method, the star is usually described as an additional point-sized particle and its motion
in the Lagrangian approach does not require special attention. Conversely, the grid-based methods 
often use cylindrical  or 
spherical coordinate systems, which are naturally centered on the star. If the barycenter 
of the star plus disk system is offset from  the center of the numerical grid, 
the star will move in response to the net non-zero gravity force exerted on it from the rest of the
system. 
However, the use of curvilinear coordinate systems with a singularity point in the coordinate
center makes it extremely difficult to calculate directly the stellar motion and its 
gravitational response on the disk and planet system.
To circumvent this problem,
a non-inertial frame of reference is often introduced, which moves with
the star (positioned in the coordinate center) in response to the net gravitational force 
of the disk and planets. The resulting acceleration
of the star is often described by the so-called indirect potential (see details in the next Section).

Grid-based numerical models dealing with the gravitational instability and fragmentation of 
protostellar disks usually neglect the stellar motion 
(see, e.g., \citealp{Pickettetal2003,Mejiaetal2005,Boleyetal2006,Boleyetal2007,BoleyDurisen2008,
Caietal2008,TsukamotoMachida2011,Boss2012,Steiman-Cameron2013,Vor2013,Lin2015}).  \citet{MichaelDurisen2010} made for the first time a focused numerical investigation of the effect of stellar motion 
induced by non-axisymmetry developed in a gravitationally unstable circumstellar disk. They 
concluded that previous findings regarding disk evolution  remain valid, while the disk gravitational instability weakens 
slightly if the stellar motion is taken into account. Their study is, however, limited to 
isolated disks and neglects the effect of the infalling envelope. Therefore, their results are  more 
applicable to protoplanetary rather than to protostellar disks, although their adopted 
disk-to-star mass ratio ($\la 0.1$) is typical for the latter.  To develop further this research, we re-investigate the effect of stellar motion on the formation and evolution of protostellar disk.

The majority of grid-based numerical investigations dealing with the gravitational interactions in 
protoplanetary disks take into account stellar motion by means of the indirect potential. 
However,  there are several works which neglect stellar motion (some examples include, but are 
not limited to \citealp{Lubowetal1999,Tanakaetal2002,Ouetal2007,Lyraetal2009a,Lyraetal2009b,
Zhuetal2011,Fungetal2014,Szulagyietal2014,Zhuetal2014,ZhuStone2014}). Therefore, it is also 
important  to explore the effect of stellar motion caused by mass asymmetry in 
protoplanetary disks.

\citet{MittalChiang2015} found that if an artificial displacement of the barycenter of the protoplanetary disk-star system is strong enough, a large-scale vortex can form via the Rossby Wave Instability (RWI, \citealp{Lovelaceetal1999}). Recently, \citet{ZhuBaruteau2016} have shown that such a displacement can be sustained by the vortex itself, forcing the star to wobble. They also found that neglecting the stellar motion significantly weakens the vortex. We note that \citet{ZhuBaruteau2016} prescribed an initially RW-unstable disk by artificially setting an axisymmetric sharp Gaussian density bump in the disk. We,
on the other hand, intend to investigate the effect of stellar motion on the formation of a RW-unstable disk and the evolution of large-scale vortices in a more self-consistent manner.

The outline of the paper is the following. In Section 2, the methodologies used for 
calculating the indirect potential in the grid-based hydrodynamic codes are presented. 
Section~3 presents the effect of stellar motions on the evolution of a protostellar disk. 
Section~4 presents the effect of stellar motion on the evolution of a protoplanetary disk hosting a massive embedded planet and a large-scale vortex formed at the viscosity transition. 
The paper ends with our conclusions on 
the importance of stellar motion in Section 5.

\section{Simulation methodology}

To investigate the effect of stellar motion on distinct evolutionary stages of accrection disks formed around young stellar objects, we employ two independent numerical
hydrodynamics codes. The reason for using two codes (FEoSaD  and GFARGO) is that they are
specifically designed to study the protostellar (early) and protoplanetary (late) stages of disk evolution, respectively. On one hand, the FEoSaD code (Formation and evolution of a star and its circumstellar disk) 
is best suited for describing the early evolution of young stellar objects, including important interconnections
and feedback mechanisms between  the central star, its disk and infalling envelope, 
such as the mass infall onto the disk, disk self-gravity, stellar irradiation, and disk radiative cooling. On the other hand, the GFARGO code, the Graphical Processor Unit (GPU) accelereated version of the FARGO code, is specifically designed for simulating the planet-disk interactions, utilizing  the Fast advection in rotating gaseous objects (FARGO) algorithm  \citep{Masset2000}. If the radial velocities are small compared to the Keplerian velocity (which is true for protoplanetary disks), the 
FARGO algorithm is about an order of magnitude faster than traditional advection schemes. 
Nevertheless, both codes 
bear some similarity as they both employ the thin-disk approximation and a similar finite-difference
solution procedure on a cylindrical numerical grid. In the following, detailed descriptions of the applied protostellar and protoplanetary disk models are given.

\subsection{Protostellar disk model}
\label{FEoSaD}
The numerical model for the formation and evolution of a star and its circumstellar disk (FEoSaD) 
is described in detail in \citet{VB2010,VB2015}
and is briefly reviewed here for the reader's convenience. Numerical simulations
start from a collapsing pre-stellar core of a certain mass, angular momentum, and temperature.
The forming star is described by the Lyon stellar evolution code \citep{Chabrier97,Baraffe2012},  
while the formation and long-term evolution of the circumstellar disk 
are described using numerical hydrodynamics simulations in the thin-disk limit. 
The evolution of the star and the disk are interconnected: the stellar evolution code takes 
the mass accretion rate onto the star provided by hydrodynamic simulations and
returns the stellar photospheric 
and accretion luminosities, which are used to calculate the disk heating rate by stellar irradiation.

The main physical processes taken into account when modeling the disk formation and evolution
include viscous and shock heating, irradiation by the forming star, 
background irradiation ($T_\mathrm{bg}=10$\,K), radiative cooling from the disk surface, and self-gravity. 
The code is written in the thin-disk limit, complemented by a calculation of the vertical 
scale height $Z$ using an assumption of local 
hydrostatic equilibrium. The resulting model has a flared structure, which guaranties that 
both the disk and envelope receive a fraction of the irradiation energy 
from the central protostar. The pertinent equations of mass, momentum, and
energy transport are

\begin{equation}
\label{cont}
\frac{{\partial \Sigma }}{{\partial t}} =  - \nabla_p  \cdot 
\left( \Sigma \bl{v}_p \right),  
\end{equation}
\begin{eqnarray}
\label{mom}
\frac{\partial}{\partial t} \left( \Sigma \bl{v}_p \right) &+& \left[ \nabla \cdot \left( \Sigma \bl{v_p}
\otimes \bl{v}_p \right) \right]_p =   - \nabla_p {\cal P}  + (\nabla \cdot \mathbf{\Pi})_p + \\ \nonumber
& + &    \Sigma \, (\bl{g}_p - \bl{g}_{\ast}), 
\label{energ}
\end{eqnarray}
\begin{equation}
\frac{\partial e}{\partial t} +\nabla_p \cdot \left( e \bl{v}_p \right) = -{\cal P} 
(\nabla_p \cdot \bl{v}_{p}) -\Lambda +\Gamma + 
\left(\nabla \bl{v}\right)_{pp^\prime}:\Pi_{pp^\prime}, 
\end{equation}
where subscripts $p$ and $p^\prime$ refer to the planar components $(r,\phi)$ 
in polar coordinates, $\Sigma$ is the mass surface density, $e$ is the internal energy per 
surface area, 
${\cal P}$ is the vertically integrated gas pressure calculated via the ideal equation of state 
as ${\cal P}=(\gamma-1) e$ with $\gamma=7/5$,
$\bl{v}_{p}=v_r \hat{\bl r}+ v_\phi \hat{\bl \phi}$ is the velocity in the
disk plane, and $\nabla_p=\hat{\bl r} \partial / \partial r + \hat{\bl \phi} r^{-1} 
\partial / \partial \phi $ is the gradient along the planar coordinates of the disk. 
The gravitational acceleration in the disk plane, $\bl{g}_{p}=g_r \hat{\bl r} +g_\phi \hat{\bl \phi}$, takes into account self-gravity of the disk, which can be calculated by solving the Poisson integral 
\citep[see details in][]{VB2010}, and the gravity of the central protostar when formed. 
Turbulent viscosity is taken into account via the viscous stress tensor 
$\mathbf{\Pi}$, which is provided in \citet{VB2010}.
We parameterize the magnitude of kinematic viscosity $\nu$ using the alpha prescription of \citet{ShakuraSunyaev1973}  
with a spatially and temporally uniform $\alpha_{\rm visc}=5\times 10^{-3}$.
The cooling and heating rates, $\Lambda$ and $\Gamma$, take the blackbody cooling and heating of the disk due to stellar and background irradiation into account 
\citep[see details in][]{VB2010}. 

To explore the effects of stellar motion, Equation~(\ref{mom}) is formulated in the non-inertial 
frame of reference moving with the star in response to the gravity force of both the disk and 
envelope. The acceleration of the star's frame of reference $\bl{g}_\ast$ can be expressed as
\begin{equation}
\bl{g}_\ast = G \int { dm(\bl{r}^\prime) \over r^{\prime 3} } \bl{r}^\prime,
\label{starAccel}
\end{equation} 
where $dm$ is the mass in a grid cell with position vector $\bl{r^\prime}$.
In practice, we find $\bl{g}_\ast$ by
first calculating its Cartesian components $g_{\ast,x}$ and $g_{\ast,y}$ as
\begin{eqnarray}
g_{\ast,x} &=&  \sum_{j,k} {F}_{j,k} \cos(\phi_{k}), \\
g_{\ast,y} &=&  \sum_{j,k} {F}_{j,k} \sin(\phi_{k}),
\end{eqnarray}
where $\phi_{k}$ is the azimuthal angle of the grid cell $(j,k)$. 
The summation is performed over all grid zones  and the force (per unit stellar mass) 
acting from the grid cell $(j,k)$ onto the star can be expressed in the following form
\begin{equation}
{F}_{j,k}= G {m_{j,k} \over r_j^2 },
\label{accel}
\end{equation}
where $m_{j,k}$ is the mass in the grid cell $(j,k)$ and $r_j$ is the radial distance to the grid cell
$(j,k)$. 

Once the Cartesian components $g_{\ast,x}$ and $g_{\ast,y}$ of the stellar acceleration are known in every grid cell, the corresponding polar-grid components can be found using the 
standard coordinate transformation formula
\begin{eqnarray}
g_{\ast,r}&=& g_{\ast,x}\cos(\phi) + g_{\ast,y} \sin(\phi),  \\
g_{\ast,\phi}&=& - g_{\ast,x}\sin(\phi) + g_{\ast,y} \cos(\phi),
\end{eqnarray}  
where $\phi$ is the azimuthal coordinate of a given grid cell.

 When the gravitational potential $\Phi$ is used in Equation~(\ref{mom}), rather than the gravitational acceleration $\bl{g}$, it is convenient to introduce the so-called indirect
potential as $\Phi_{\rm ind}=\bl{r} \cdot \bl{g}_\ast$. 
The gradient of $\Phi_{\rm ind}$ equals $\bl{g}_\ast$ because the latter depends on $r^\prime$ rather than on \bl{r} (see Equation~\ref{starAccel}). The last term in the right hand side of Equation~(\ref{mom}) can then
be substituted with the following term: $-\Sigma \nabla (\Phi + \Phi_{\rm ind}),$ and the indirect potential
in a grid cell $(j,k)$ can be calculated as:
\begin{equation}
\Phi_{\rm ind}^{j,k} =    r_{j} \cos(\phi_k) g_{\ast,x} + r_{j} \sin(\phi_k) g_{\ast,y} ,
\end{equation}
where summation is performed over all grid zones. In this paper, we use 
$-\nabla (\Phi + \Phi_{\rm ind})$
instead of the acceleration $\bl{g}-\bl{g}_\ast$ in Equation~(\ref{mom}).

Equations (\ref{cont})-(\ref{energ}) are solved using the method of finite differences
with a time-explicit, operator-split solution procedure
in polar coordinates ($r,\phi$) on a numerical grid with 
$512\times512$ grid zones. The radial grid is logarithmically spaced, while the azimuthal grid is 
equispaced. Advection is treated using the third-order piecewise
parabolic interpolation scheme \citep{CW1984}. The update of the internal energy per surface
area due to cooling and heating is done implicitly using
the Newton-Raphson method of root finding, complemented by
the bisection method where the Newton-Raphson iterations fail
to converge.

To avoid too small time steps, we introduce a ``sink cell'' at $r_{\rm sc}=6.0$~au and 
impose a free outflow inner boundary condition so that the matter is allowed to flow from the 
inner active computational region
into the sink cell, but is prevented from flowing in the opposite direction. 
At the outer computational boundary a free inflow-outflow boundary condition is imposed,
so that the matter is allowed to flow in and out of the active computational domain.
This is done to prevent the artificial accumulation of matter near the outer
computational  boundary in the non-inertial frame of reference (but see discussion in Section~\ref{init}).
More details on the code and the
solution procedure can be found in \citet{VB2010,VB2015}.
The parameters of the initial model for studying the evolution of protostellar disks (model~1)
are given in Section~\ref{init}. 

\subsection{Protoplanetary disk model}
\label{sec_GFARGO}

The effect of stellar motion in a protoplanetary disk can be important if the latter has a large-scale
asymmetry. The sources of these asymmetries can be eccentric gaps opened by a giant planet 
\citep{KleyDirksen2006,Regalyetal2010},  global disk eccentricities excited due to the gravitational perturbation of a massive perturber \citep{Lubow1991,Kleyetal2008} or a giant planet \citep{Regalyetal2014}, and large-scale vortices excited by the Rossby Wave Instability triggered by the gap-opening 
planet \citep{deVal-Borroetal2007} or by the viscosity transition \citep{Lietal2005}. 
Migration of planets can also be affected by stellar motion as it is governed by the 
angular momentum exchange between the planet and the disk.

 We investigate the gap and disk eccentricity formation caused by an embedded non-migrating giant planet
on a circular orbit (models~2 and 3), the type II migration of high-mass planets (models~4 and 5), and
the formation of a large-scale vortex in the vicinity of a sharp viscosity transition (model~6), 
where the outer edge of the so-called dead zone is assumed to be located. To this end,
we employ the two-dimensional locally isothermal hydrodynamical simulations with and without taking
the effect of the indirect potential term into account. 

For the hydrodynamical modeling we use the GPU based version of the FARGO code \citep{Masset2000}, which solves the vertically integrated continuity and Navier-Stokes equations (Eqs. (1) and (2)) on the cylindrical coordinate system using a locally isothermal approximation. In our simulations, the star is located at the center of the numerical domain. The indirect potential has two sources: 1) the indirect potential of the star due to the shift of the barycenter by the disk material
\begin{equation}
        \Phi^{j,k}_{\rm ind,disk}=\sum_{j',k'}\left(\frac{m_{j',k'}}{r_{j',k'}^3}[{x_{j',k'}x_{j,k}+y_{j',k'}y_{j,k}}]\right)
        \label{eq:indt-disc}
\end{equation}
similar to what is described in Sect.\,2.2.; and 2) the indirect potential caused by the shift of the barycenter due to a massive planet
\begin{equation}
        \Phi^{j,k}_\mathrm{ind,pl}=M_\mathrm{p}\left(\frac{1}{R_{j,k}^3}[x_{j,k}x_\mathrm{p}+y_{j,k}y_
        \mathrm{p}]\right).
\label{eq:indt-pl}
\end{equation}
In Equations (\ref{eq:indt-disc}) and (\ref{eq:indt-pl})
\begin{equation}
        r_{j,k}=\sqrt{x_{j,k}^2+y_{j,k}^2},
\end{equation}
\begin{equation}
        R_{j,k}=\sqrt{x_\mathrm{p}^2+y_\mathrm{p}^2}
\end{equation}
$m_{j,k}$ and  $x_{j,k}$, $y_{j,k}$, are the mass and Cartesian coordinates of the grid cell $j,k$, while $M_\mathrm{p}$ and $x_\mathrm{p}$, $y_\mathrm{p}$ are the mass and Cartesian coordinates of
the planet, respectively. In simulations which do not take the effect of the indirect potential into account, we simply neglect the calculation of $\Phi^{j,k}_{\rm ind,disk}$ and $\Phi^{j,k}_\mathrm{ind,pl}$.

In the local isothermal approximation, the radial temperature profile is $T(R) \sim R^{-1}$, and the equation of state of the gas, ${\cal P}(R,\phi)=c_\mathrm{s}(R)^2\Sigma(R,\phi)$, depends only on the density and the local sound speed $c_\mathrm{s}(R)=\Omega(R)H(R)$, where $H(R)$ is the local pressure scale height. We note that both $c_\mathrm{s}(R)$ and $H(R)$ remain constant in time due to the locally isothermal assumption. The disk is assumed to be flat, for which case $H(R)=hR$, where $h$ is the disk aspect ratio, which is set to $h=0.05$ for all models. The disk's viscosity is approximated by the $\alpha$-prescription of \citet{ShakuraSunyaev1973} with  $5\times10^{-5}\leq\alpha_\mathrm{visc}\leq5\times10^{-3}$. 

We adopt 1~au for the unit of length and $1.0~M_\odot$ for the unit of mass. The unit of time is such that the orbital period of a planet at 1\,au is $2\pi$.  With this scaling the gravitational constant equal unity. For models~2-5 we use a frame that co-rotates with the planet.

We use logarithmically distributed  $N_R=256$ radial and $N_\phi=512$ equidistant azimuthal grid cells for  all protoplanetary disk models. To model the eccentricity excitation of the inner disk, the relatively close inner boundary is required, thus our computational domain extends from  $R_\mathrm{in}=0.3$\, au to $R_\mathrm{out}=30$\,au in models $2-5$ (see, e.g., \citealp{Regalyetal2014}). For model~6, we use a slightly larger disk inner and outer domain boundary, that is, $R_\mathrm{in}=3$~au and  
$R_\mathrm{out}=50$~au. The larger inner boundary is required in order to model the long-term evolution of large-scale vortices. The larger outer disk boundary is necessary to avoid artificial boundary effects due to the vicinity of the disk's outer edge to the large-scale vortex.

To take the disk thickness into account, we use $\epsilon H(R_\mathrm{p})$ as the smoothing of the gravitational potential of the planet with $\epsilon=0.6$ in protoplanetary disk simulations following the receipt in \citet{Kleyetal2012} and \citet{Mulleretal2012}. In the planet-bearing disk models, a circumplanetary disk forms around the giant planet, which affects its migration history (models~4 and 5). To circumvent this effect we remove a certain amount of the torque exerted on the planet by the disk  according to \citet{Cridaetal2009} in models~4 and 5. Thus, the force exerted on the planet by the material inside the planetary Hill sphere is multiplied by $1-\exp(-(d/R_\mathrm{H})^2)$, where $R_\mathrm{H}=a_\mathrm{p}(q/3)^{1/3}$ is the planetary Hill radius and $d$ is the distance between the given grid cell center and the planet.

To speed-up the simulations, the damping boundary condition \citep{deVal-Borroetal2006} is applied at  the inner boundary. The damping criteria at the inner domain edge is set such that the waves are killed close to the inner boundary $R<1.2R_\mathrm{in}$. With this setting, the time step (which is calculated according to the Courant-Friedrich-Levi criteria described in \citealp{Masset2000}) remains considerably large because sharp density enhancements (i.e., shock waves) cannot appear at the boundary. The outflow boundary is used at the disk's outer edge in all protoplanetary disk models.

\section{Early disk evolution (protostellar disk)}
In this section, we investigate the effect of stellar motion on the early 
evolution of protostellar disks, when they are still actively accreting material
from the parental cores. For this purpose, we use the FEoSaD code described
in Section~\ref{FEoSaD}.  We start our numerical simulations from the gravitational collapse of a starless cloud core, continue into the embedded phase of star formation, during which
a star, disk, and envelope are formed, and terminate our simulations when the age of the
star exceeds 0.4~Myr. 

\subsection{Initial conditions}
\label{init}
For the initial conditions, we consider a gravitationally unstable pre-stellar core submerged
into a low-density environment, both described by the following radial gas surface density
distribution 
\begin{equation}
\label{dens}
\Sigma = \left\{
\begin{array}{ll}
{r_0 \Sigma_0 \over \sqrt{r^2+r_0^2}} \,\,\,\,\, 
       \mathrm{for}~r\le R_{\rm core}, \nonumber \\
\Sigma_{\rm ext} \hskip 0.8cm \mathrm{otherwise},            
\end{array} \right. 
\end{equation}
where $\Sigma_0$ is the gas surface
density at the center of the core and $r_0 =\sqrt{A} c_{\rm s}^2/\pi G \Sigma_0 $
the radius of the central plateau, $c_{\rm s}$ the initial sound speed in the core,
$R_{\rm core}$ the radius of the core, and $\Sigma_{\rm ext}$ the gas surface density in 
the external environment. The gas surface density distribution of the core can
be obtained (to within a factor of unity) by integrating the 
3D gas density distribution characteristic of 
Bonnor-Ebert spheres with a positive density-perturbation amplitude $A$ \citep{Dapp09}.
The value of $A$ is set to 1.2. The core has a fixed ratio $R_{\rm core}/r_0=6.0$, 
implying that it is initially unstable to gravitational collapse. The initial
temperature is set everywhere throughout the core to 10~K. For the chosen
core radius of $R_{\rm core}=0.09$~pc and central surface density of 
$\Sigma_0=2\times 10^{-2}$~g~cm$^{-2}$, the resulting initial core mass is $M_{\rm core}=1.18$~$M_\odot$.
The total computational domain extends to 0.18~pc and,  for the value of 
$\Sigma_{\rm ext}=6.4\times 10^{-5}$~g~cm$^{-2}$, the total mass
contained in the external environment is $M_{\rm ext}=0.02~M_\odot$, which is only
about 1\% that of the core mass.

The initial angular velocity is described by the following equation 
\begin{equation}
\Omega=2\Omega_0 \left( {r_0\over r}\right)^2 \left[\sqrt{1+\left({r\over r_0}\right)^2
} -1\right],
\label{omega}
\end{equation}
as expected 
for slowly contracting pre-stellar cores with conservation of angular momentum \citep{Basu1997}.
Here, $\Omega_0=0.49$~km~s$^{-1}$~pc$^{-1}$ is the angular velocity in the center of the core. 
For the chosen parameters of our model, the ratio of rotational to gravitational energy is
$\beta=5.5\times 10^{-3}$. We hereafter refer to this
model as model~1.

The reason for adding an external constant-density environment
to the pre-stellar core (rather than just considering an isolated core) 
is that the shape of the core near its outer edge can be artificially
distorted in the non-inertial, accelerated frame of reference.
We employed the logarithmically spaced grid in the radial direction, 
which resulted in a good numerical resolution in the inner regions where the 
protostellar disk forms and evolves, but in a poor numerical resolution near the outer boundary.
The imperfections in the 
implementation of the outer computational boundary lead to
artificial accumulation or depression in the gas density near the boundary when
the material in the computational domain is accelerated due to the indirect potential. 
This artificial distortion of the core shape often has an undesirable feedback 
effect on the disk dynamics. We experimented with
different outer boundary conditions (free boundary, extrapolated density and velocity profiles),
but the result was not satisfactory. 

The development of special numerical schemes that 
allow for the moving boundaries (so that the flow of material 
through the outer boundary caused by the acceleration of the coordinate system is minimized)  
could potentially help to alleviate the problem. Our temporary solution to 
this problem was to extend the outer boundary sufficiently far from the disk and fill in
the rest of the numerical grid with a low-density gas so that its gravitational potential
would have little effect on the dynamics of the system, even if this external environment becomes distorted
due to the boundary effects. As we show later in the paper,
this helps to mitigate the problem for the initial several hundred thousand years of protostellar
disk evolution. However, because
of the pressure disbalance between the core and the external environment,
the outer regions of the core gradually expand (while the inner part continues collapsing onto the
star) and ultimately reach the outer computational boundary.
At this time instance, we have to stop the simulations.

\subsection{The effect of stellar motion on the protostellar disk and envelope}

\begin{figure}
 \centering
  \resizebox{\hsize}{!}{\includegraphics{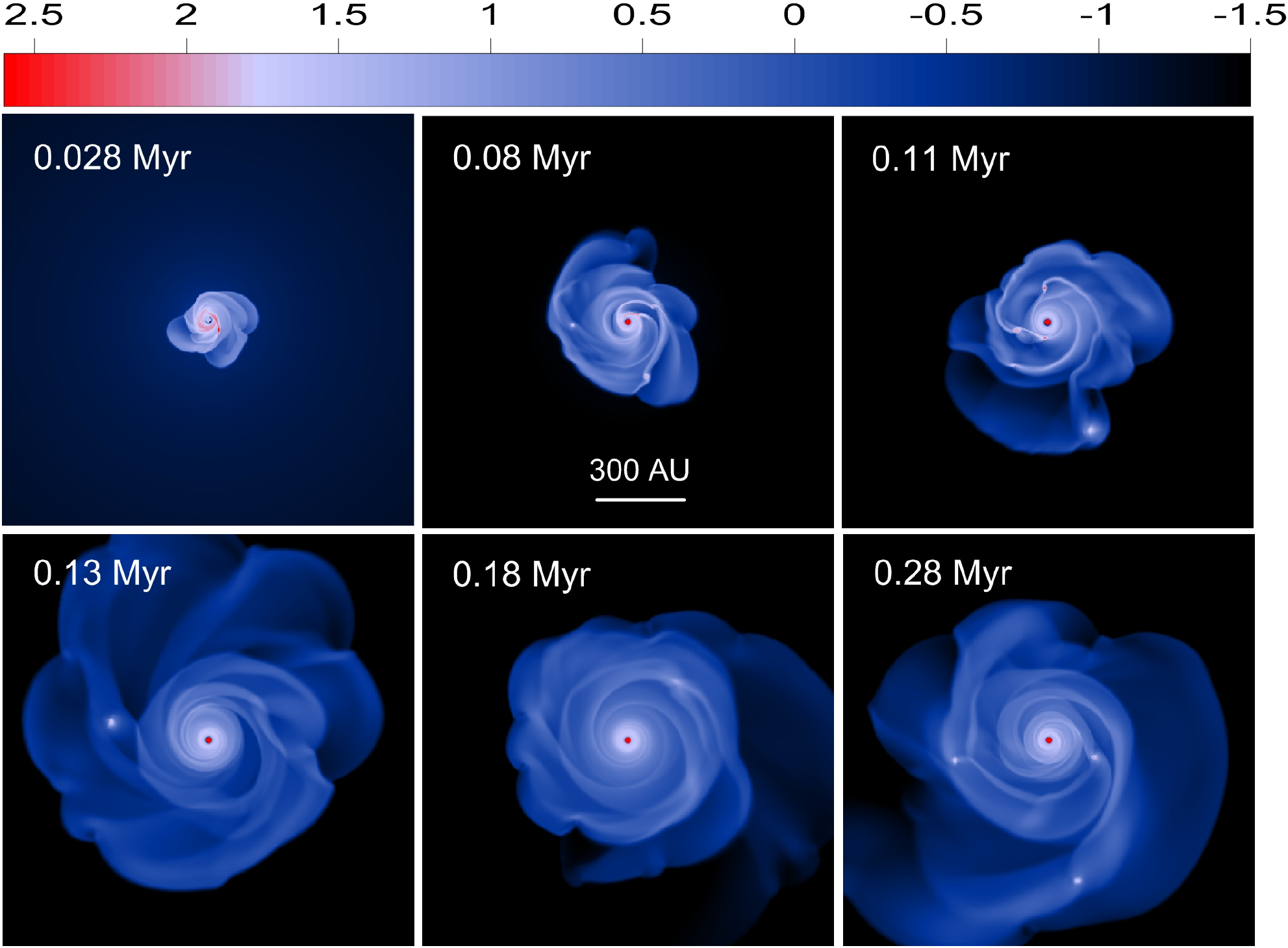}}
  \caption{Gas surface density distribution in model~1 in the inner $1400\times1400$~au box focusing on the disk. Six evolution
  times counted from the formation of the star are shown. The scale bar is in g~cm$^{-2}$ (log units).}
  \label{fig1}
\end{figure}

In this section, we focus on possible
differences in the properties of the disk and infalling envelope 
that may arise when  stellar motion is taken into account.
We differentiate between models with and without stellar motion by adding the plus sign to the corresponding
model number, that is, model~1+ means that the stellar motion is taken into account, 
while model~1 means
the opposite. The chosen initial parameters of the pre-stellar core ($M_{\rm core}=1.18~M_\odot$ 
and $\beta=5.5\times 10^{-3}$)
imply that the core collapse will
lead to the formation of a gravitationally unstable disk \citep{Vor2013}.

Figure~\ref{fig1} presents the gas surface density in model~1 
in the inner $1400\times1400$~au$^2$ box at several
time instances after the formation of the central star. We note that the total computational 
domain is much greater, but for now we focus only on the innermost regions occupied by the disk. 
Gravitational instability develops in the disk soon after its formation due to the continuing
mass loading from the parental core. The spiral structure has a complicated pattern and dense
gaseous clumps form within the spiral arms. The vigorous gravitational instability and
fragmentation continues for at least 0.3~Myr until the gas reservoir in the parental core exhausts
and the infalling envelope (that continuously replenishes the disk material lost 
via accretion onto the star) dissipates.

\begin{figure}
 \centering
  \resizebox{\hsize}{!}{\includegraphics{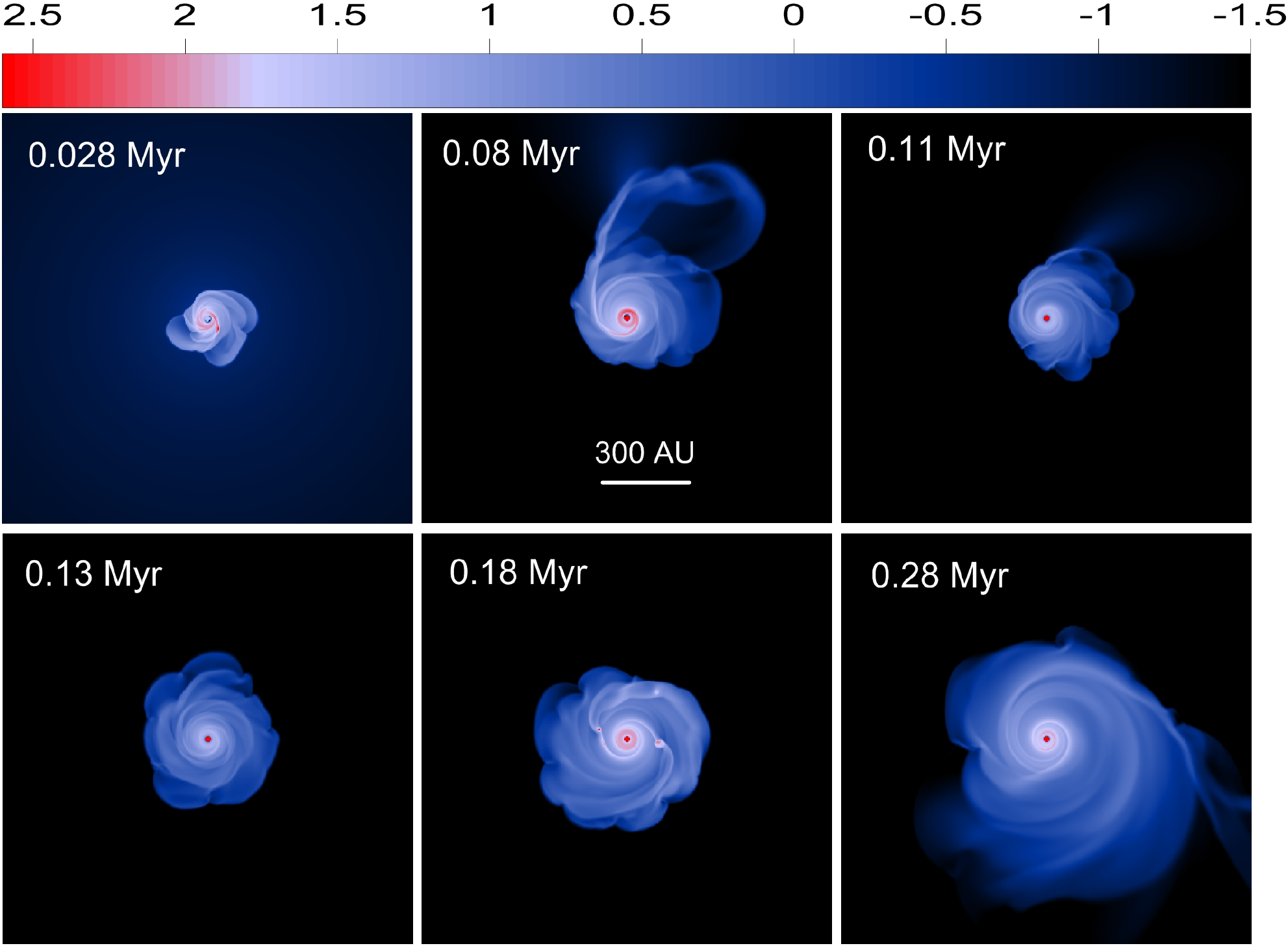}}
  \caption{Similar to Figure~\ref{fig1}, but for model~1+ for which case stellar motion is included.}
  \label{fig2}
\end{figure}

Figure~\ref{fig2} shows the gas surface density in model~1+ at the same 
time instances as in model~1. The stellar motion is introduced $2\times10^{4}$~yr after 
the instance of disk formation\footnote{Before the disk formation,
the collapse proceeds axisymmetrically because of the chosen initial axisymmetric shape of the 
pre-stellar core, so that there is no necessity to introduce the stellar motion.},  or 
$2.8\times 10^{4}$~yr after the formation of the central star; the corresponding
gas surface density distribution is shown in the upper-left panel in Figure~\ref{fig2}.
This delay is introduced to let the disk grow and acquire some material. 
Otherwise, we found that the disk becomes significantly
distorted in the initial formation stage, leading to the termination of numerical simulations. 
This problem is likely to be related to the imperfections of the inner computational  boundary
(near to which the disk naturally forms) and requires further investigation.

\begin{figure}
 \centering
  \resizebox{\hsize}{!}{\includegraphics{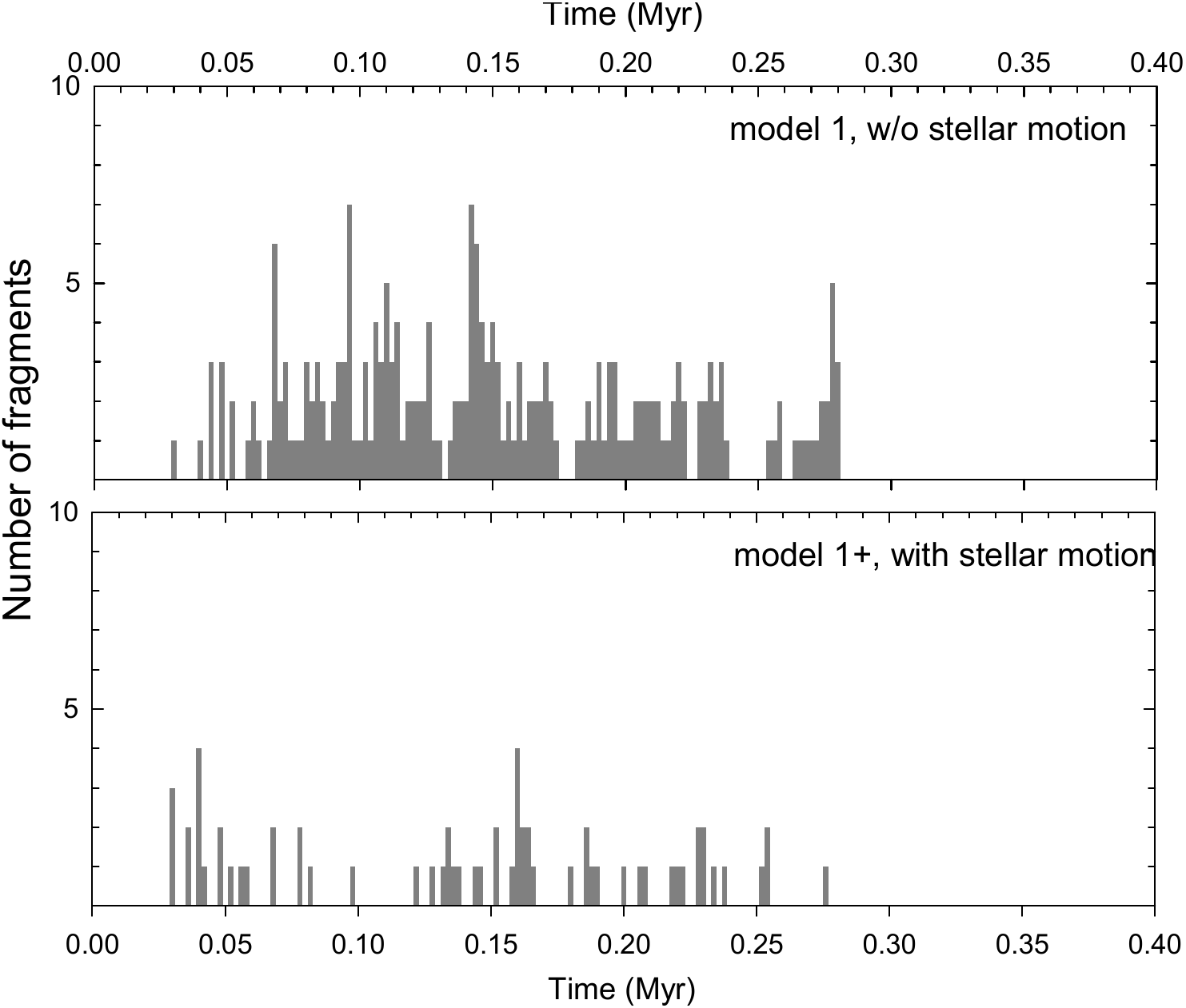}}
  \caption{Number of fragments in the disk at a given time elapsed since the formation of the star.
  The top and bottom panels correspond to models~1 and 1+ (without and with stellar motion, respectively). }
  \label{fig3}
\end{figure}

A visual inspection of Figures~\ref{fig1} and \ref{fig2} indicates that the stellar motion has 
a notable effect on the disk radius and the strength of  gravitational instability: the disk radius is smaller and the disk fragmentation is less frequent in model~1+ than in model~1.
To quantify these differences, we calculated the number of fragments in the disk every 2000~yr
using the fragment tracking mechanism explained in detail in \citet{Vor2013}. This algorithm 
requires that the fragment be pressure-supported, having a negative
pressure gradient with respect to the center of the fragment, and that the fragment 
be kept together by gravity, having the deepest potential well at the center of the
fragment. The results are shown in Figure~\ref{fig3} in the top panel (model~1) and bottom
panel (model~1+). While model~1 forms fragments almost continuously starting from $t=0.03$~Myr 
until $t=0.27$~Myr,
model~1+ exhibits only sporadic fragmentation episodes. The maximum number of the fragments
in the disk at any given time is also greater in model~1 ($N_{\rm fr,max}=7$) than in model~1+ 
($N_{\rm fr,max}=4$).

\begin{figure}
 \centering
  \resizebox{\hsize}{!}{\includegraphics{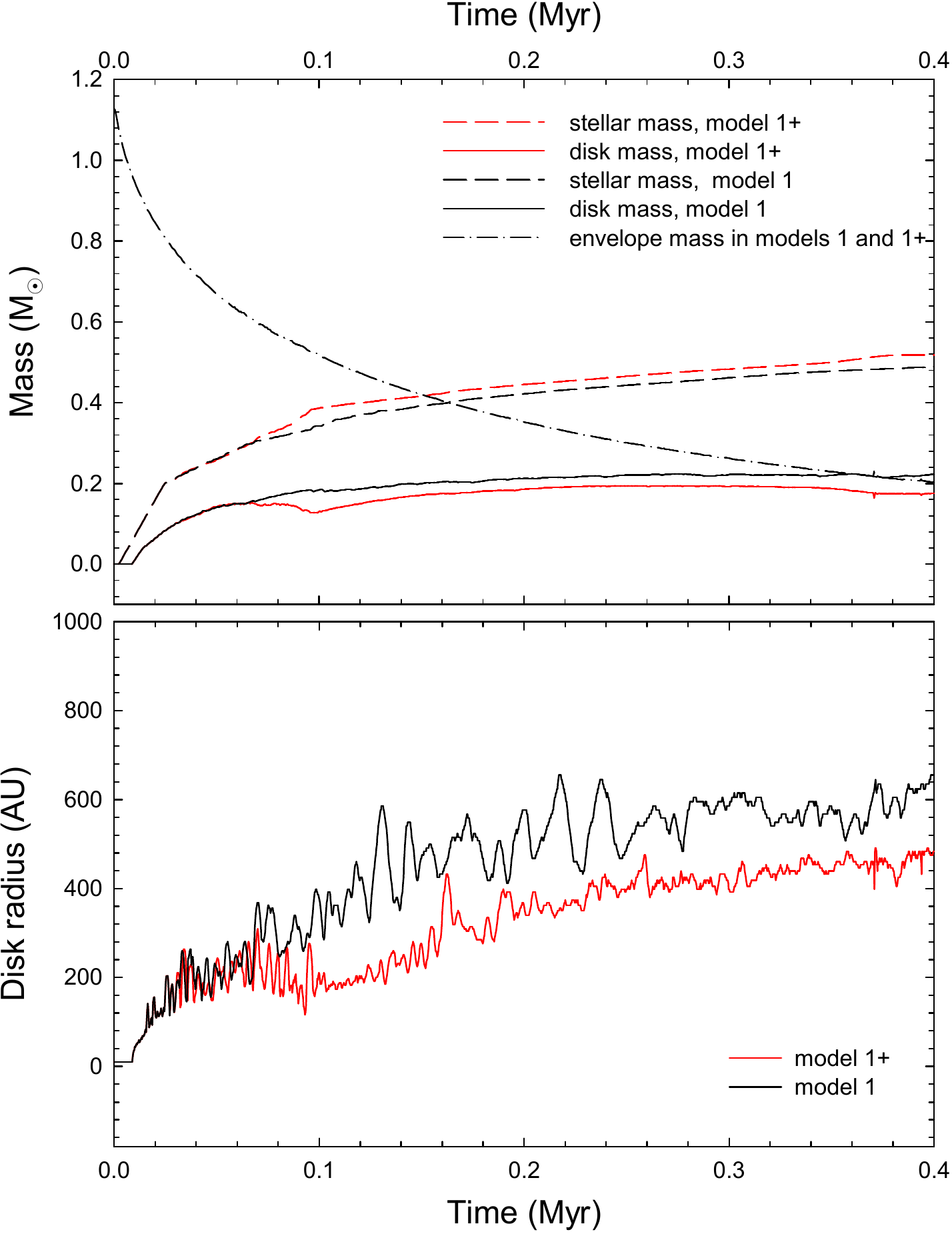}}
  \caption{ Top panel: Stellar, disk, and envelope masses in model~1 (black-colored lines) and model~1+ (red-colored lines). The envelope mass is the same in both models. Bottom panel: The disk radius in
model 1 (black-solid line) and  model~1+ (red solid line).}
  \label{fig4}
\end{figure}

To understand the reason for this difference, we plot the total disk masses, stellar
masses, and disk radii as a function of time in Figure~\ref{fig4}. In particular, the red and black lines show the data for
models with and without stellar motion, respectively. The black dash-dotted lines correspond
to the envelope mass, which is similar in both models. Evidently, model~1+ has a somewhat smaller and less massive disk than model~1. We think this is caused by the fact that, in the model 
with stellar motions, part of the total angular momentum of the collapsing core goes into the orbital
motion of the star around the center of mass of the whole system. On the other hand, 
in the model without stellar motion, all angular momentum of the collapsing core 
is transferred to the disk. As a result, the disk in model~1 has more angular momentum and a greater support against 
gravity than in model~1+, helping the former to sustain a more massive and extended disk. 
This conjecture  is supported by direct calculation of the specific angular
momentum of the disk: it is equal to $J_{\rm disk}=2.4\times 10^{25}$~cm$^2$~s$^{-1}$ in model~1+
and $J_{\rm disk}=3.3\times 10^{25}$~cm$^2$~s$^{-1}$ in model~1.

The spiral structure in Figures~\ref{fig1} and \ref{fig2} is complicated and does not
show any regular, long-lasting pattern. This is typical for unstable protostellar disks subject to
fragmentation caused by the continuous mass-loading from parental cores 
\citep[e.g.,][]{VB2010,Tsukamoto2013}. 
Nevertheless, it is possible to analyze the spiral pattern by calculating the global Fourier 
amplitudes defined as
\begin{equation}
C_{\rm m} (t) = {1 \over M_{\rm disk}} \left| \int_0^{2 \pi} 
\int_{r_{\rm sc}}^{R_{\rm disk}} 
\Sigma(r,\phi,t) \, e^{im\phi} r \, dr\,  d\phi \right|,
\label{fourier}
\end{equation}
where $M_{\rm disk}$ is the disk mass and $R_{\rm disk}$ is the disk's physical 
outer radius.  When the disk
surface density is axisymmetric, the amplitudes of all modes
are equal to zero. When, say, $C_{\rm m}(t) = 0.1$, the perturbation
amplitude of spiral density waves in the disk is 10\% that of the
underlying axisymmetric density distribution.

\begin{figure}
 \centering
  \resizebox{\hsize}{!}{\includegraphics{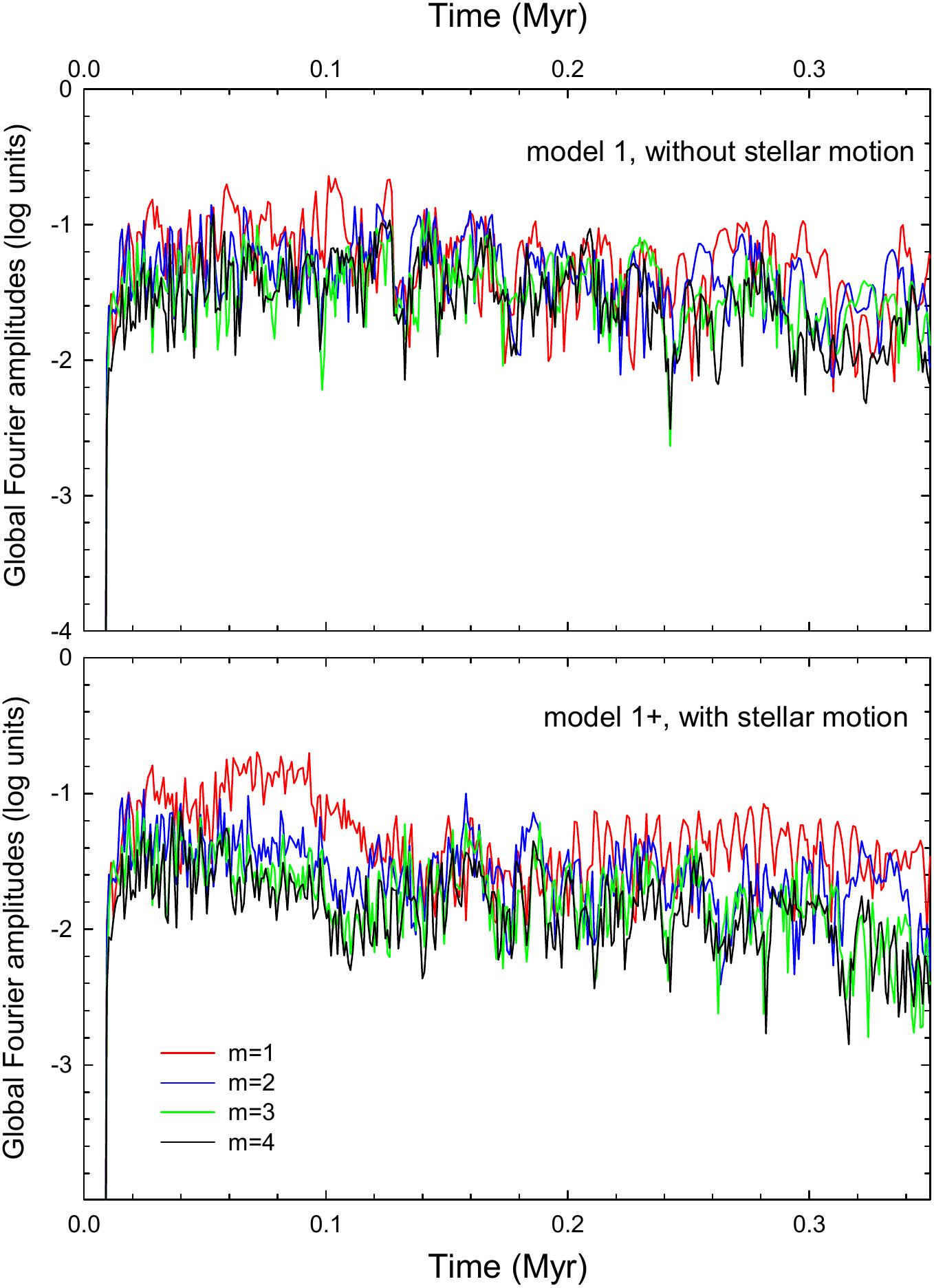}}
  \caption{Global Fourier amplitudes of the disk as a function of time in model~1 (top panel)
  and in model~1+ (bottom panel) calculated for the first four spiral modes.}
  \label{fig6}
\end{figure}

Figure~\ref{fig6} presents the time evolution of the global Fourier amplitudes for the
first four modes in model~1+ (top panel) and model~1 (bottom panel).  The $m=1$ spiral mode clearly dominates in model~1+, except for a short time period around $t=0.14-0.2$~Myr, when all the four modes
have similar amplitudes. In model~1, on the other hand, the dominance of the $m=1$ mode is only marginal.
At the later times, the amplitudes in model~1+ are somewhat smaller  than in model~1,
which is consistent with a lower disk mass in model~1+. The dominance of the
$m=1$ mode in the model with stellar motion may be related to the SLING 
(stimulation by the long-range interaction of Newtonian gravity) effect discussed in, for example, 
\citet{Shu1990}. We note that \citet{MichaelDurisen2010} found no SLING effect in their numerical
hydrodynamics simulations with stellar motion, 
possibly due to the fact that they considered an isolated disk with a smaller disk-to-star mass ratio.

\begin{figure}
 \centering
    \resizebox{\hsize}{!}{\includegraphics{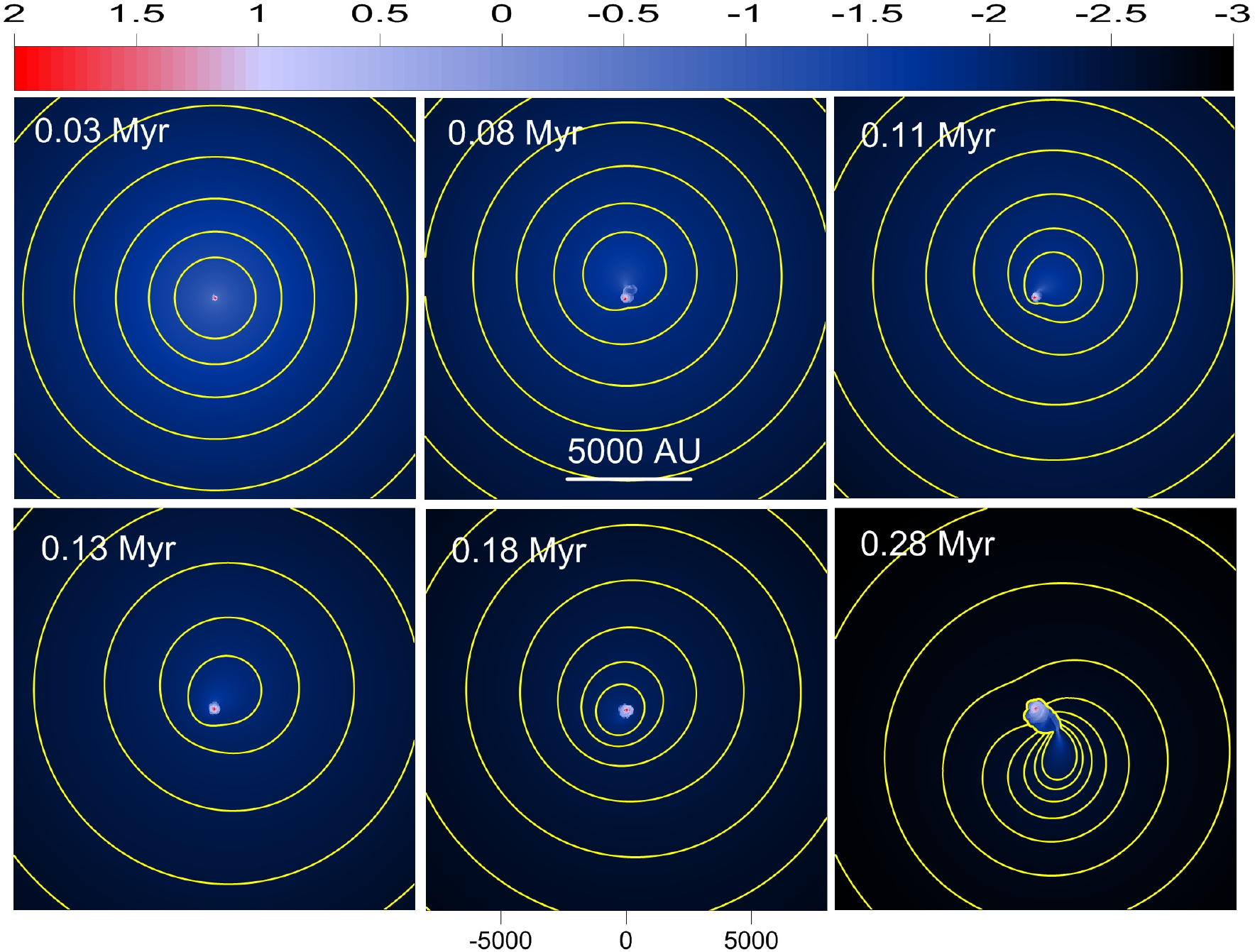}}
  \caption{Gas surface density distribution in the inner $16000\times16000$~au$^2$ box covering
  both the disk and the infalling envelope in model~1+ 
  at six times after the formation of the central star. The yellow iso-surface density
  contour lines are added to highlight
  the inner envelope structure. The scale bar is in log g~cm$^{-2}$.}
  \label{fig5}
\end{figure}

Stellar motion has a notable effect not only on the disk, but also on the structure
of the infalling envelope -- the remnant of the parental pre-stellar core. Figure~\ref{fig5} 
presents a zoomed-out view of the computational domain,
this time covering the $16000\times16000$~au$^2$ area which includes both the disk and infalling envelope.
The disk occupies the inner several hundred au, while the rest is occupied by the envelope.
The iso-surface density contour lines are added to the gas surface density maps to better 
characterize the density structure of the envelope.
At the very early stage ($t=0.03$~Myr), just after the stellar motion has been turned on, the envelope
has an axisymmetric shape consistent with the initial configuration. However, already at $t=0.08$~Myr
the deviation of the inner envelope structure from a purely axisymmetric shape becomes evident. 
This is because the stellar motion has a feedback effect not only on the disk, but also on the 
envelope, at least on its inner part.  
However, after $t\approx0.2$~Myr the expansion wave caused
by the pressure imbalance at the interface between the core and the external environment reaches
the outer computational boundary and the boundary effects discussed in Section~\ref{init} start to artificially
distort the shape of the whole envelope,   as is seen in the $t=0.28$~Myr panel. 
At this stage, we terminate the simulations.

\begin{figure}
 \centering
  \resizebox{\hsize}{!}{\includegraphics{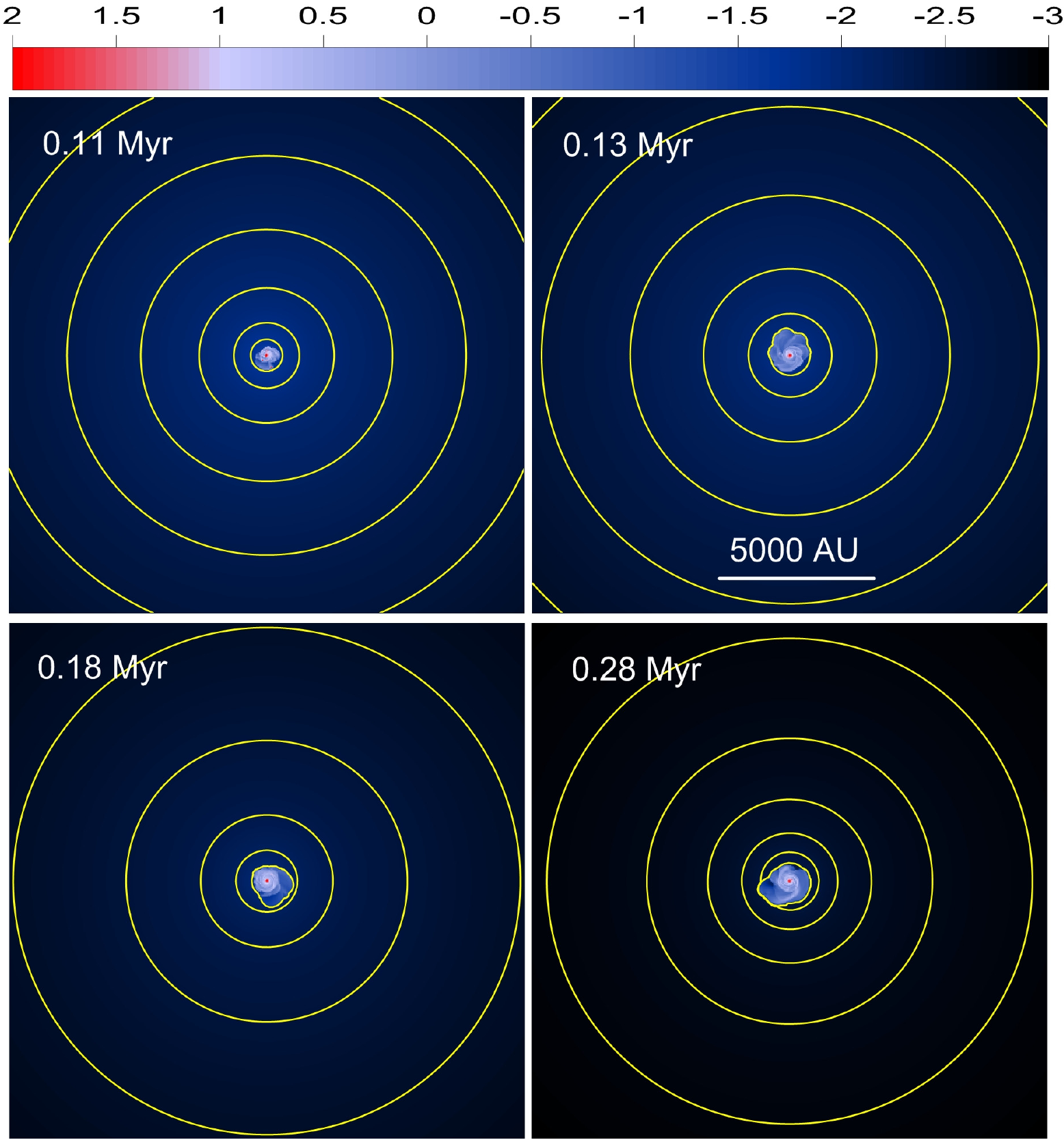}}
  \caption{Similar to Figure~\ref{fig5}, only for model~1 without stellar motion. Only four evolutionary
  times are shown.}
  \label{fig5a}
\end{figure}

To check if the deviation of the inner envelope from a purely axisymmetric configuration 
is indeed caused by stellar motion, we plot in Figure~\ref{fig5a}
the gas surface density distribution in model~1 (without stellar motion) for the 
inner $8000\times8000$~au$^2$ box.  The yellow iso-surface density contour lines highlight the envelope structure. 
Evidently, the envelope is axisymmetric, except for very inner regions where it meets with the
non-axisymmetric disk. 

\begin{figure}
 \centering
  \resizebox{\hsize}{!}{\includegraphics{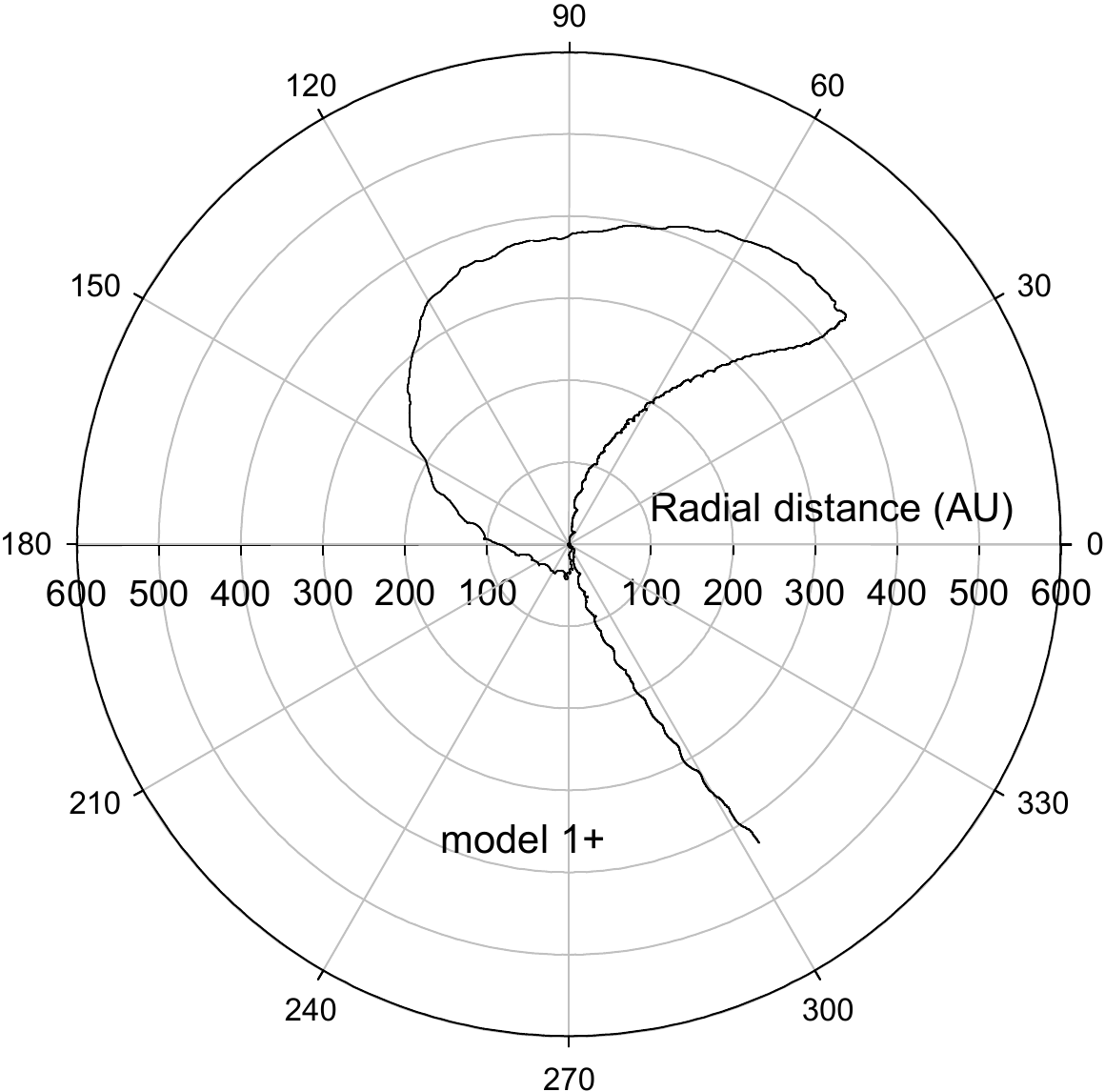}}
  \caption{Position of the center of mass in model~1+ at the polar coordinate system. The center of
  mass is initially at ($r=0,\phi=0$), but shows significant excursions during the subsequent evolution
  due to the non-axisymmetric structure of the infalling envelope. Small-scale wobbling in response 
  to the non-axisymmetric disk is also evident. }
  \label{fig7}
\end{figure}

Figure~\ref{fig7} gives the position of the center of mass (CoM) of the whole system 
(star+disk+envelope+external environment) on the polar grid ($r,\phi$). 
The  CoM shows large excursions from the initial position ($r=0, \phi=0$), as well as small-scale wobbling. The large excursions are the result of the gradual 
transformation in the structure of 
the infalling envelope from a purely axisymmetric to largely non-axisymmetric shape,
while the small wobbling is the reaction to the non-axisymmetric gravitational perturbation of the disk.
Since the mass contained in the envelope is significant 
(it is always greater than the disk mass and is also greater than
the stellar mass in the early evolution phase, see Figure~\ref{fig4})  
the effect of the non-axisymmetric inner envelope on the stellar motion is 
stronger than that of the disk. The spiral arms and fragments in the disk often tend to cancel 
the effect of each other on the star,
which also acts to reduce the stellar wobbling due to the disk.

Finally, we note that in our adopted thin-disk approximation all the disk and envelope 
mass is concentrated in the midplane. The immediate consequence of this assumption is that
any non-axisymmetry in the disk or envelope has a stronger effect on the star as compared
to the real 3D case. In the 2D case, the gravitational force has only 
two plane components,
while in the 3D case there is also a vertical component. The latter would contribute to the
vertical excursions of the star (rather than radial ones due to the plane components) and
would vanish in systems with equatorial 
symmetry.  This problem can be partly solved in the 2D case by splitting the disk vertical column into $i$ layers of equal width and calculating
the input from every vertical layer individually. The resulting force acting upon the star from
the grid cell $(j,k)$ can than be expressed as
\begin{equation}
\tilde{F}(i,j,k)= 2 G \sum_i {m_{i,j,k} \over (r_j^2 + z_{i}^2)^{3/2}} r_j ,
\label{accelnew}
\end{equation}
where the vertical height of each layer
is denoted by $z_i$ and the summation is performed over all vertical layers. 
The mass of each layer 
$m_{i,j,k}$  can be found by assuming a constant or Gaussian distribution of the gas volume density within the vertical column. A factor of two expresses the fact that the disk has an equatorial symmetry.
Our test calculations indeed indicate that the resulting envelope structure is less distorted and 
the problem of the outer computational boundary becomes less severe, though not solved completely.
From a general point of view, we note that the gravitational instability in 
the 3D disks is expected to be weaker than in the 2D ones, which may also reduce 
the gravitational effect of the disk  upon the star. Other 3D effects, 
such as disk warps, may reduce the radial excursions of the star, but introduce 
notable vertical motions.

\subsection{Protostellar accretion rates}

\begin{figure}
 \centering
  \resizebox{\hsize}{!}{\includegraphics{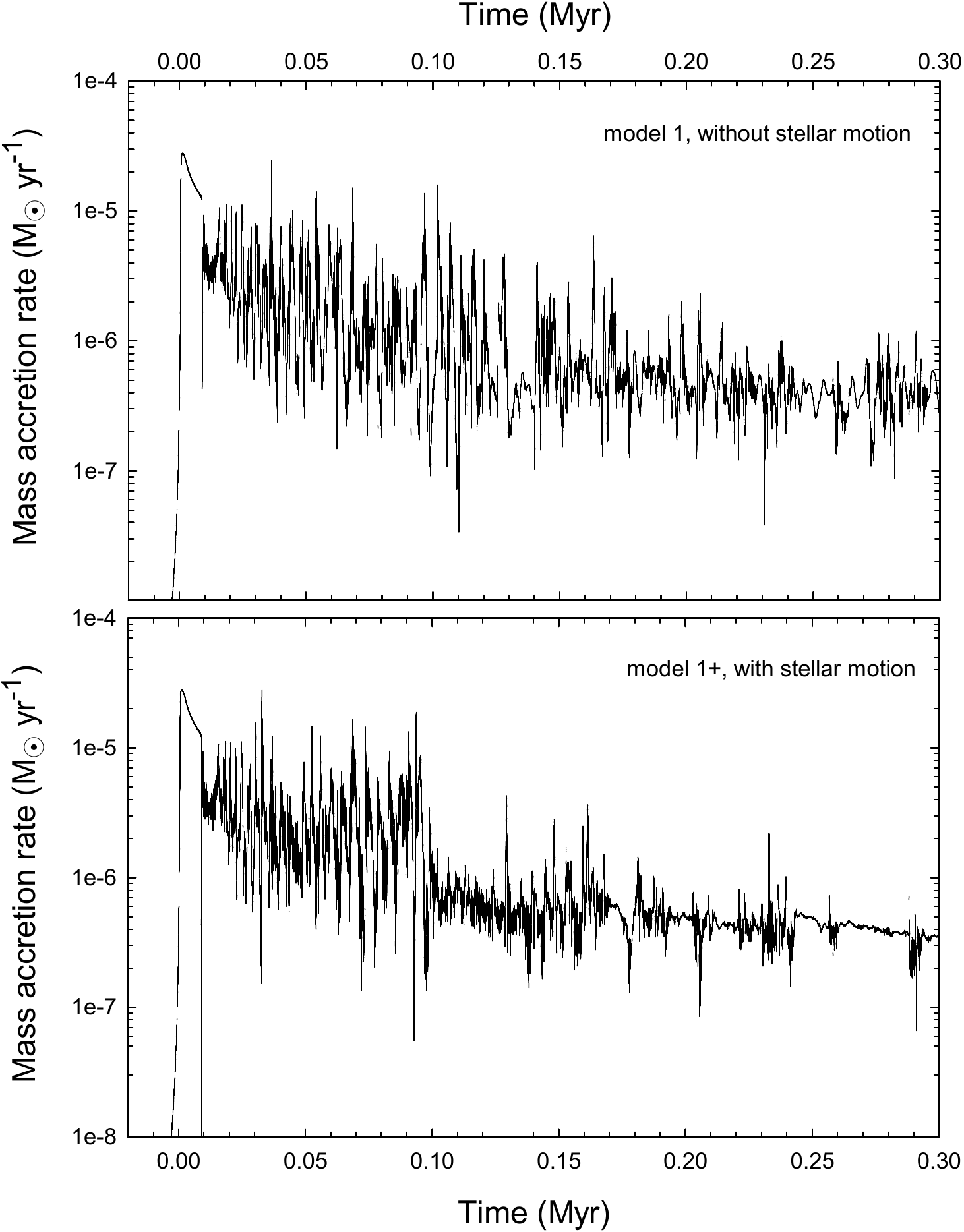}}
  \caption{Mass accretion rates as a function of time in model~1 (top panel) and model~1+ (bottom panel). }
  \label{fig8}
\end{figure}

In this section, we investigate the effect of stellar motion on the 
mass accretion rate defined as the mass passing through the inner computational boundary
per hydrodynamical time step, $\dot{M}=-2\pi r_{\rm sc} \Sigma v_{\rm r}$.  
Previous numerical simulations without stellar motion demonstrated
that $\dot{M}$ often exhibits time-variable behavior with episodic bursts,
if the disk mass is sufficient to trigger gravitational instability and fragmentation  
\citep[e.g.,][]{VB2006,VB2010,Machida2011,VB2015}. This accretion pattern 
is part of the emerging episodic accretion paradigm, wherein the mass accretion 
onto the star is not constant or gradually declining in time, as predicted by the classic 
spherical collapse models \citep[e.g.,][]{Shu1977}, but is time-variable with strong 
accretion bursts separated
by longer periods of quiescent, low-rate accretion \citep[see, e.g., a recent review by][]{Audard2014}.
It is therefore important to assess the effect of stellar motion on the mass accretion rate.

Figure~\ref{fig8} presents the mass accretion rates as a function of
time in models~1+ (top panel) and model~1 (bottom panel).
 In the early evolution $t\la 0.1$~Myr, the accretion rate 
in models with and without stellar motion exhibits a similar behavior:
it quickly rises to a peak value at $t=0$~Myr (the instance of protostar formation), declines 
gradually for a short period of time  ($\approx0.01$~Myr), during which the matter is directly
accreted from the infalling envelope onto the protostar, and then develops 
order-of-magnitude variations when the disk around the protostar forms and gains mass.   
Accretion bursts amounting to $\dot{M}=\mathrm{a~few} \times 10^{-5}~M_\odot$~yr$^{-1}$ are evident
in both models.

In the later evolution ($t\ga 0.1$~Myr), the mass accretion rate in model~1+ is characterized 
by a somewhat reduced variability in comparison to model~1,
though still showing several order-of-magnitude accretion bursts. This reduction in the amplitude
of accretion variations is consistent with a lower disk mass in model~1+. Overall, the stellar motion
somewhat reduces the accretion variability, particularly in the later evolution stages, mainly due to
a reduced disk mass. This reduction can, however, be offset by 
a small increase in the initial core mass and/or angular momentum 
(which would result in the formation of a more massive and gravitationally unstable disk).

\section{The late disk evolution (protoplanetary disk)}

In this section, we investigate how stellar motion affects the evolution of protoplanetary disks perturbed by a massive planet, by the migration of planets, and by the formation of large-scale vortices in protoplanetary disks. We assume that the protoplanetary disk has already been formed and an embedded high-mass planet revolves on a circular orbit (models~2-5). In simulations dealing with the formation of vortices at the dead zone outer edge, we impose a sharp viscosity transition (model~7). Similar to the previous section, we run protoplanetary disk simulations with and without stellar motion. Models in which we take the effect of the indirect potential into account are labeled with an additional plus sign.

The simulations span $4\times10^4$\,yrs (corresponding to about 3500 orbits at 5\,au) in models~2-5. To study the vortex formation, much longer simulations are required. Therefore, we  run simulations up to $2\times10^5$\,yrs in model~6, corresponding to about 4811 orbits at 12\,au -- a distance 
where the large-scale vortex develops.

\subsection{Initial conditions}
\label{sec:ppdisk_init}

\begin{figure*}
  \includegraphics[width=\columnwidth]{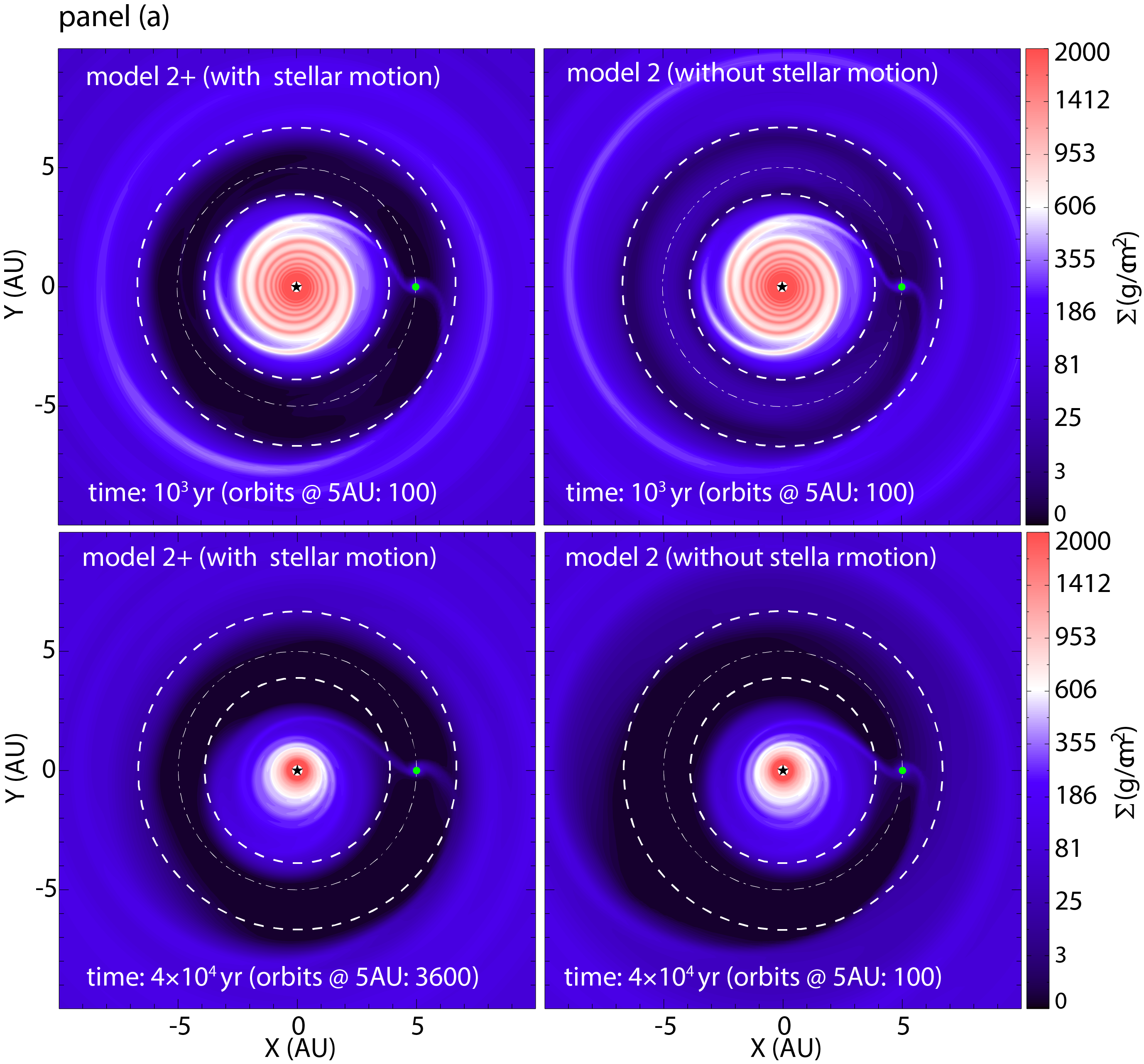}
  \includegraphics[width=\columnwidth]{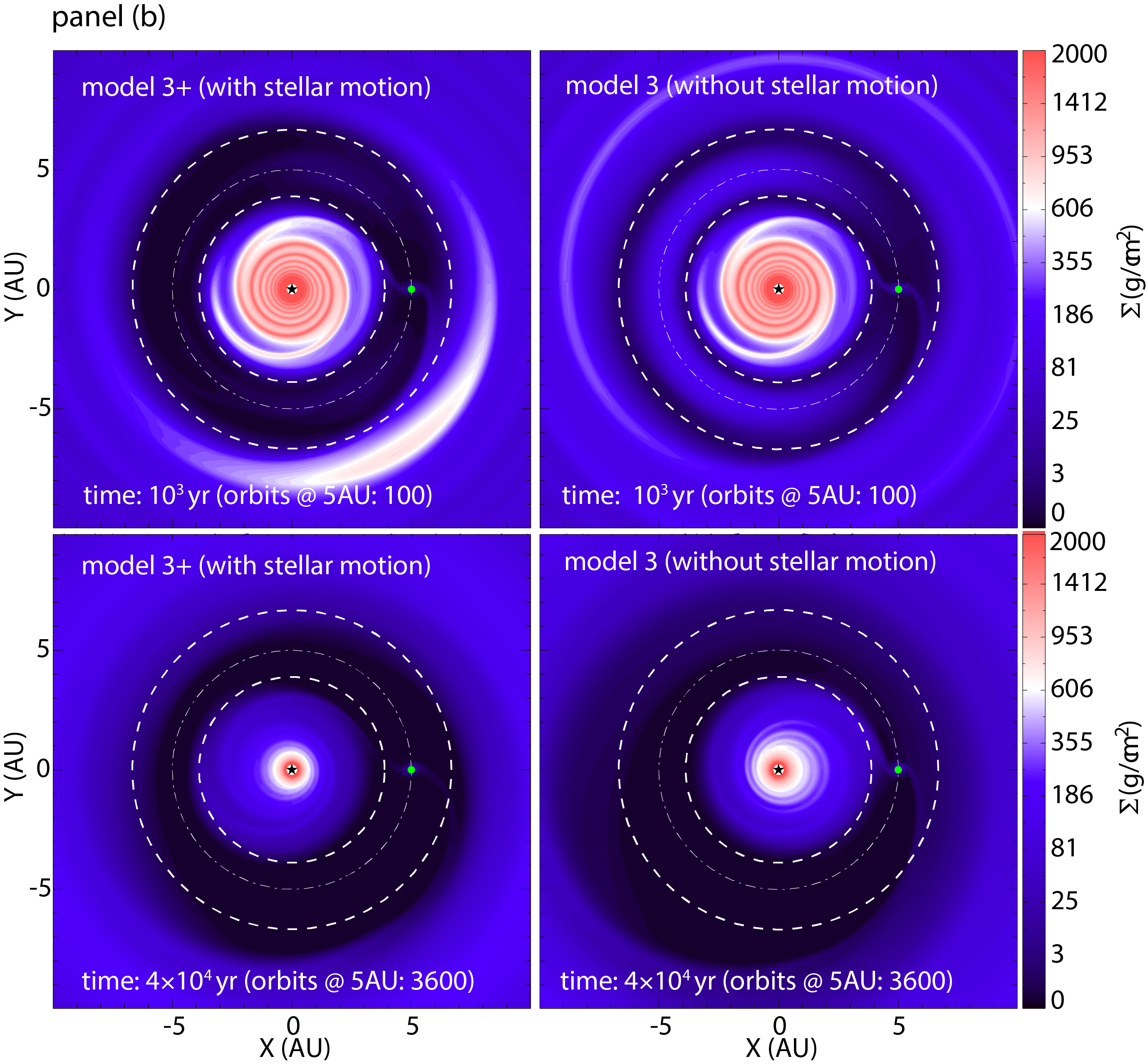}
  
  \caption{Density snapshots in models~2 and 3 for which case a non-migrating giant planet is embedded in a disk with  $\alpha_\mathrm{visc}=5\times10^{-3}$ (panel a) and $5\times10^{-5}$ (panel b). In each panel the left and right columns correspond to simulations in which the indirect potential is taken into account (model~2+ and 3+) and neglected (model~2 and 3), respectively. Snapshots are taken after the gap is opened at $t=10^3$\,yr corresponding to about 100'th orbits (upper row), while at the end of simulation at $t=4\times10^4$\,yr corresponding to about 3600'th orbits lower row).}
  \label{fig:model-3-4}
\end{figure*}

The initial surface density profile of the gas has the following form: $\Sigma(R,\phi)=\Sigma_0 R^{-p}$, where $p$ is set to 1.0 and $\Sigma_0$ is set to $1.06815\times10^{-4}$, which corresponds to a disk 
of $0.02~M_\odot$ inside the computational domain. Since the Toomre Q parameter \citep{Toomre1964} is $hM_*/\pi R^2\Sigma(R,\phi)>1$ throughout the disk in all models, it seems plausible to neglect the disk self-gravity.\footnote{In models~2-3 the planet is fixed at 5\,au, where $Q\simeq30$. In models~4-5 the planet migrates inwards, thus  $Q\geq30$ is always true at its orbital distance. In model~6 initially $Q\simeq12$ at the viscosity transition, while later $Q$ decreases to about 3 as gas accumulates there.} We note, however, that the disk self-gravity affects the migration rate of planets; it slightly slows down the migration (see, e.g., \citealp{Cridaetal2009}) even for low-mass disks where self-gravity is weak. The vortex formation at the planetary gap edge is also known to be affected by the disk self-gravity, which shifts the most unstable modes to higher mode numbers, or even suppresses the RWI for 
sufficiently high disk masses with $Q\leq2.5$ (see, e.g., \citealp{MamatsashviliRice2009,LinPapaloizou2011}). Recently, \citet{ZhuBaruteau2016} showed that including self-gravity results in significantly weakened large-scale vortices if $Q\leq\pi/(2h)$. Here we want to investigate the effect of the indirect term only, which requires long-term simulations. Therefore, the computationally-demanding self-gravity 
calculation is neglected for simplicity.

In models~2 and 3, we study the disk response to a non-migrating high-mass planet ($q=4.7\times10^{-3}$, corresponding to $M_\mathrm{p}=5\,M_\mathrm{Jup}$ for a Solar mass star). In model~2 the viscosity parameter is set to a canonical value of $\alpha=5\times10^{-3}$. Model~3 is a nearly inviscid version of model~2 with $\alpha=5\times10^{-5}$. With these settings, we expect the formation of various types of disk asymmetry, e.g., a local disk eccentricity at the gap edge (see, e.g., \citealp{KleyDirksen2006,DAngeloetal2006,Regalyetal2010}), global disk eccentricity inside the planetary orbit (see, e.g., \citealp{Regalyetal2012,Regalyetal2014}), and large-scale vortex formation at the outer gap edge (see, e.g., \citealp{Regalyetal2012}). We note that the low viscosity assumption is necessary to maintain a long-lasting vortex at the planetary gap edges as was shown earlier by \citet{deVal-Borroetal2007,Ataieeetal2013,Fuetal2014}. 

In models~4 and 5, the migration of high-mass planets ($q=9.54\times10^{-4}$ and $2.38\times10^{-3}$, corresponding to $M_\mathrm{p}=1.0$ and $2.5\,M_\mathrm{Jup}$) is investigated. With these planetary masses,  no excitation of disk eccentricity is observed, for which case the migration history might be more complicated. The planets are freely migrating in the disk (after the gap is fully fledged), having a canonical viscosity of $\alpha=5\times10^{-3}$. With these models, we investigate the effect of stellar motion on the migration history of high-mass planets.

In models~2-5, initially the planet orbits at $a_\mathrm{p}=5$\,au and is not allowed to accrete gas, that is, its mass is constant. For non-migrating cases (models~2 and 3), the planetary orbit is kept fixed. For migrating planets  (models~4 and 5), the planet is allowed to feel the disk potential after the gap is fully fledged.
 
To model the formation and evolution of the large-scale vortex developed due to the RWI in model~6, we introduce a sharp viscosity transition at the dead zone outer edge. We assume that the disk has an accretionally inactive dead zone, where the value of $\alpha$ is smoothly reduced such that $\alpha(R)=\alpha\delta_\alpha(R)$. The viscosity reduction is given by 
\begin{equation}
        \label{eq:deltaalpha}
        \delta_\alpha(R)=1-\frac{1}{2}\left(1-\alpha_\mathrm{mod}\right)\left[1-\tanh\left(\frac{R-R_\mathrm{dze}}{\Delta R_\mathrm{dze}}\right)\right],
\end{equation}
where $\alpha_\mathrm{mod}=0.01$ is the depth of the turbulent viscosity reduction. To quantify the radius of viscosity reduction, $R_\mathrm{dze}=12$\,au is used, where we adopt the results of \citet{MatsumuraPudritz2005}, who found that $R_\mathrm{dze}$ lies between 12\,au and 36\,au, depending on the density of the disk. We assume $\Delta R_\mathrm{dze}=1H_\mathrm{dze}$, where $H_\mathrm{dze}=R_\mathrm{dze}h$ is the disk scale-height at the viscosity reduction, which corresponds to $\Delta R_\mathrm{dze}=0.6$\,au. We note that the total width of the viscosity transition given by Equation\,(\ref{eq:deltaalpha}) is about $2\Delta R_\mathrm{dze}$, which corresponds to 1.2\,au.

\subsection{Disk response to a giant planet (models~2 and 3)}

We investigate the effect of stellar motion on the disk response to the gravitational perturbation of a giant planet embedded in the disk in models~2 and 3. The viscous gap opening criterium is satisfied (i.e., $M_\mathrm{p}\gg40\alpha c_\mathrm{s}^2H/\Omega_\mathrm{p}a_\mathrm{p}^2$  ) in models~2 (viscous) and 3 (nearly inviscid) for a five Jupiter mass planet \citep{LinPapaloizou1993}. As a result, a gap forms after a couple of orbits, independent of whether stellar motion is taken into account or not. 

\begin{figure}
  \resizebox{\hsize}{!}{\includegraphics{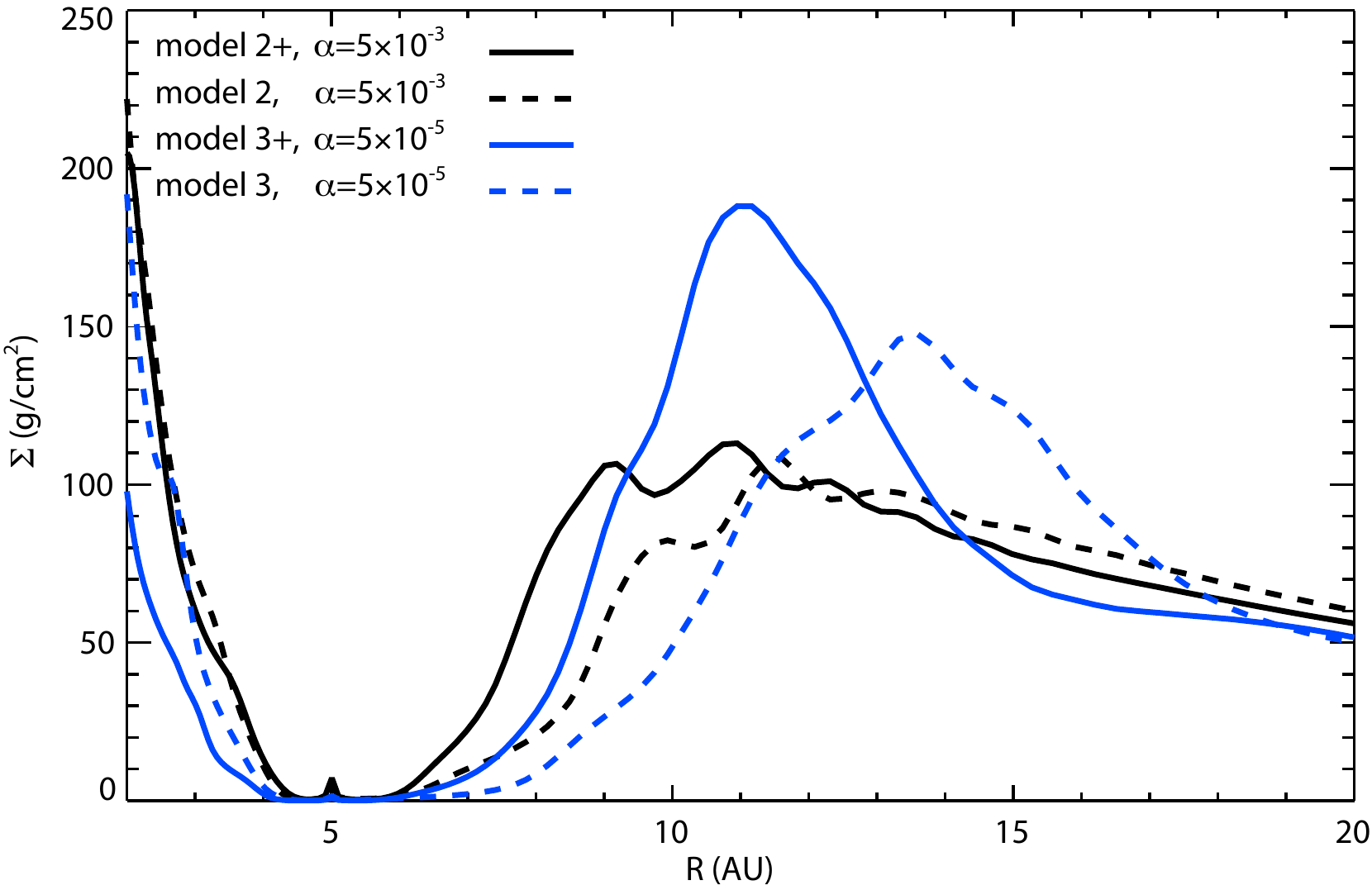}}
  \caption{Azimuthally averaged radial density profiles in models~2 and 3 at the end of simulations (corresponding to the lower row of Figure\,\ref{fig:model-3-4}). Solid and dashed lines correspond to models in which the indirect potential is taken into account and neglected, respectively.}
  \label{fig:model-3-4-densprofiles}
\end{figure}

The upper and lower rows of Figure\,\ref{fig:model-3-4} show the gap structure at $t=10^3$\,yr and at the end of simulations $t=4\times10^4$\,yr, respectively.  In the early phases, the depletion of the horseshoe region is significantly slower for models without indirect potential (compare upper rows in panels (a) and (b) of Figure\,\ref{fig:model-3-4}). In the later phases, the gap width is larger in models where the stellar motion is neglected. From the azimuthally averaged radial density profiles (Figure\,\ref{fig:model-3-4-densprofiles}) it is clearly visible that  the inner gap edge forms nearly at similar distance, while the outer gap edge is always at larger distances in models without stellar motion.

Another notable difference can be identified regarding the large-scale vortex formation at the gap edges. The large-scale vortex develops due to the excitation of the RWI at the gap edges thanks to the development of a sharp density gradient \citep{Lietal2005,Lyraetal2009b,Regalyetal2012}. However, due to the viscous evolution of the gap edge, the density jumps smear out and the gap edge is not RW-unstable any more. As a result, the vortex dissolves. We observed this phenomenon regardless of whether the stellar motion
is taken into account or not. We emphasize that the density gradient at the outer gap edge smears out faster in model~2 without stellar motion. This explains why no vortex is present after ten orbits 
in the high viscosity models~2+ and 2 (upper figures in panel (a) of Figure\,\ref{fig:model-3-4}) and is present only in model~3+ (upper figures in panel (b) of Figure\,\ref{fig:model-3-4}). We note, however, that the vortex lifetime is observed to be shorter than 100 orbits even in the low-viscosity regime, independent of whether the stellar motion is on or off.

\begin{figure}
  \resizebox{\hsize}{!}{\includegraphics{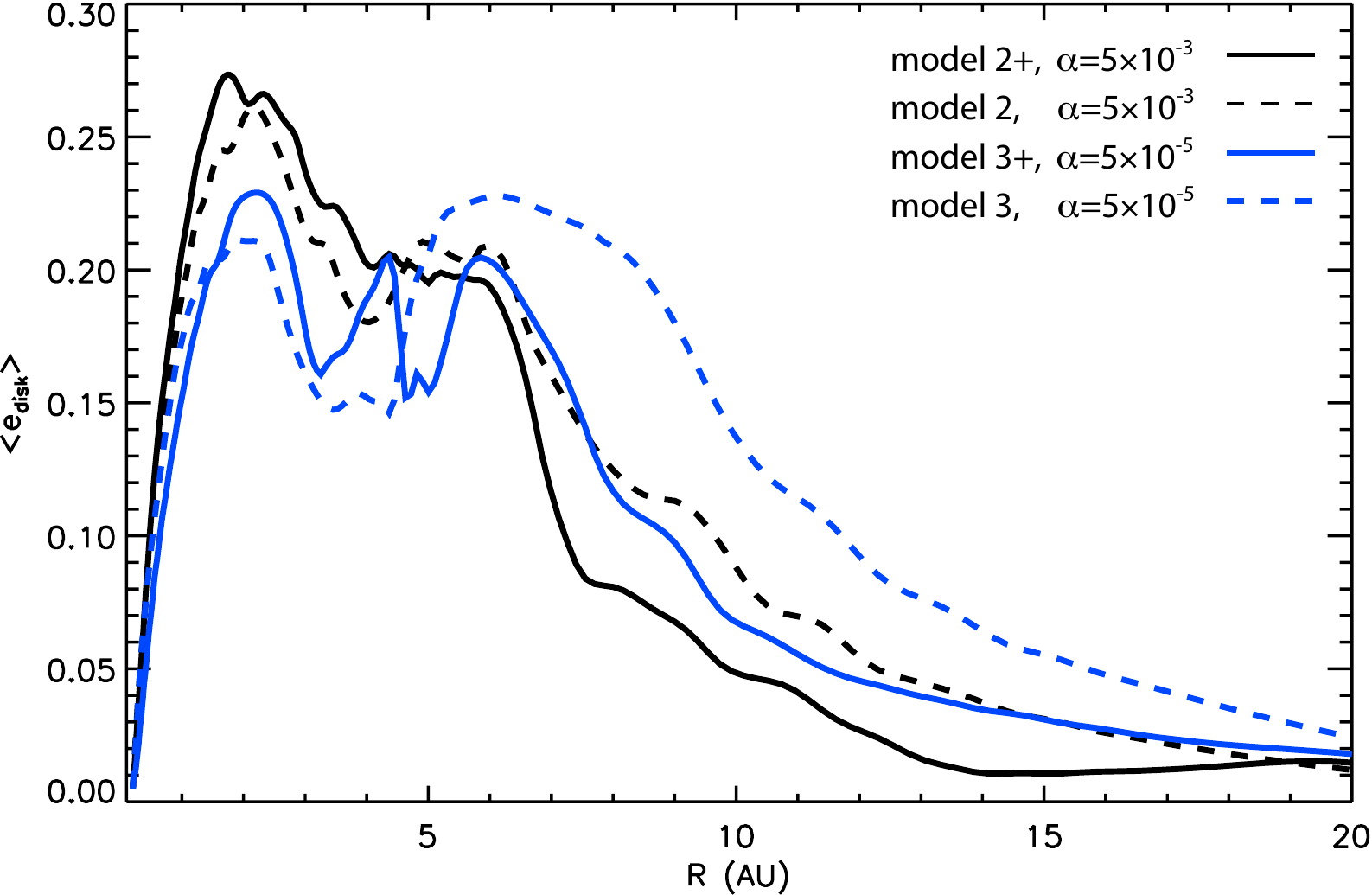}}
  \caption{Azimuthally averaged eccentricity profile of the disk at the end of the simulations  in models~2 and 3. Solid and dashed lines correspond to models in which the indirect potential is taken into account and neglected, respectively.}
  \label{fig:model-3-4-eccprofiles}
\end{figure}

As can be seen in Figure\,\ref{fig:model-3-4}, a significant level of disk eccentricity develops by the end of the simulations in both models. Not only the gap edge but also the inner disk become 
globally eccentric, as was shown in \citet{Regalyetal2014}. One can calculate the disk eccentricity at each grid cell using Eqs.\,(5)-(7) of \citet{Regalyetal2010}. After azimuthal averaging, this 
results in the radial eccentricity profile of the disk shown in Figure\,\ref{fig:model-3-4-eccprofiles}. Evidently, the eccentricity in the inner disk (with respect to the planetary orbit)  is nearly independent whether we take the stellar motion into account or not. However, the outer disk eccentricity, especially close to the gap outer edge, tends to be larger for models without stellar motion. Since the gap width is inferred from the azimuthally averaged radial density profile (Figure\,\ref{fig:model-3-4-densprofiles}), the somewhat larger disk eccentricity (apparent in the vicinity of the gap) can explain the larger gap.

\begin{figure}
  \resizebox{\hsize}{!}{\includegraphics{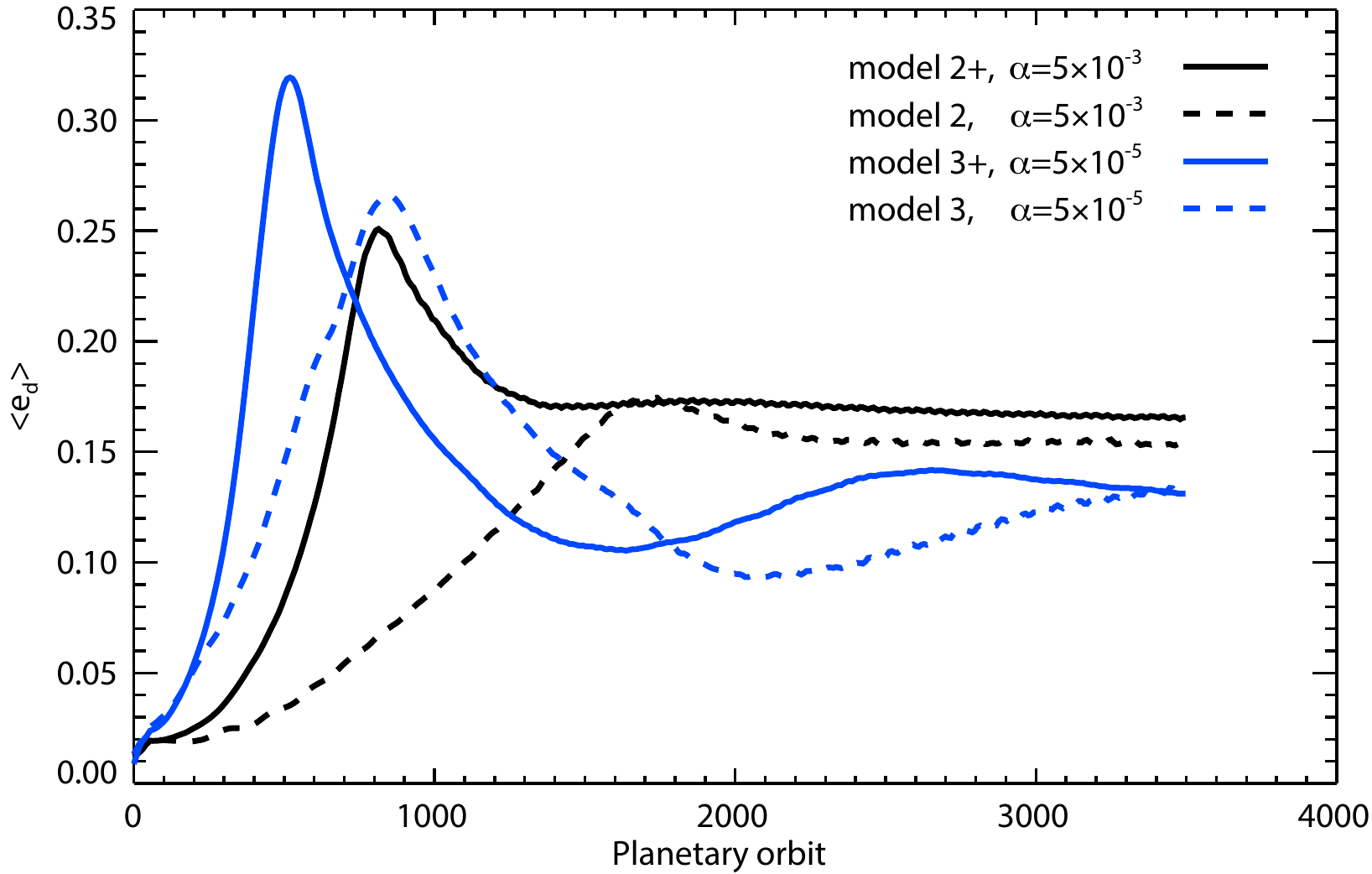}}
  \caption{Evolution of averaged disk eccentricity  in models~2 and 3. Solid and dashed lines correspond to models in which the indirect potential is taken into account and neglected, respectively.}
  \label{fig:model-3-4-eccgrowth}
\end{figure}

Radial averaging of the eccentricity profiles results in a sole number which represents the disk global eccentricity. It is clearly visible on the plot of the disk global eccentricity as a function of time (Figure\,\ref{fig:model-3-4-eccgrowth}) that a quasi-static eccentric disk state with $e_\mathrm{disk}\simeq0.15$ is developed by the end of the simulations, independent of whether the stellar motion is on or off. However, the eccentricity growth time-scale is smaller for models with stellar motion 
(models~2+ and 3+).

\subsection{Migration of giant planets (models~4 and 5)}

\subsubsection{On migration timescales}
\begin{figure}
  \resizebox{\hsize}{!}{\includegraphics{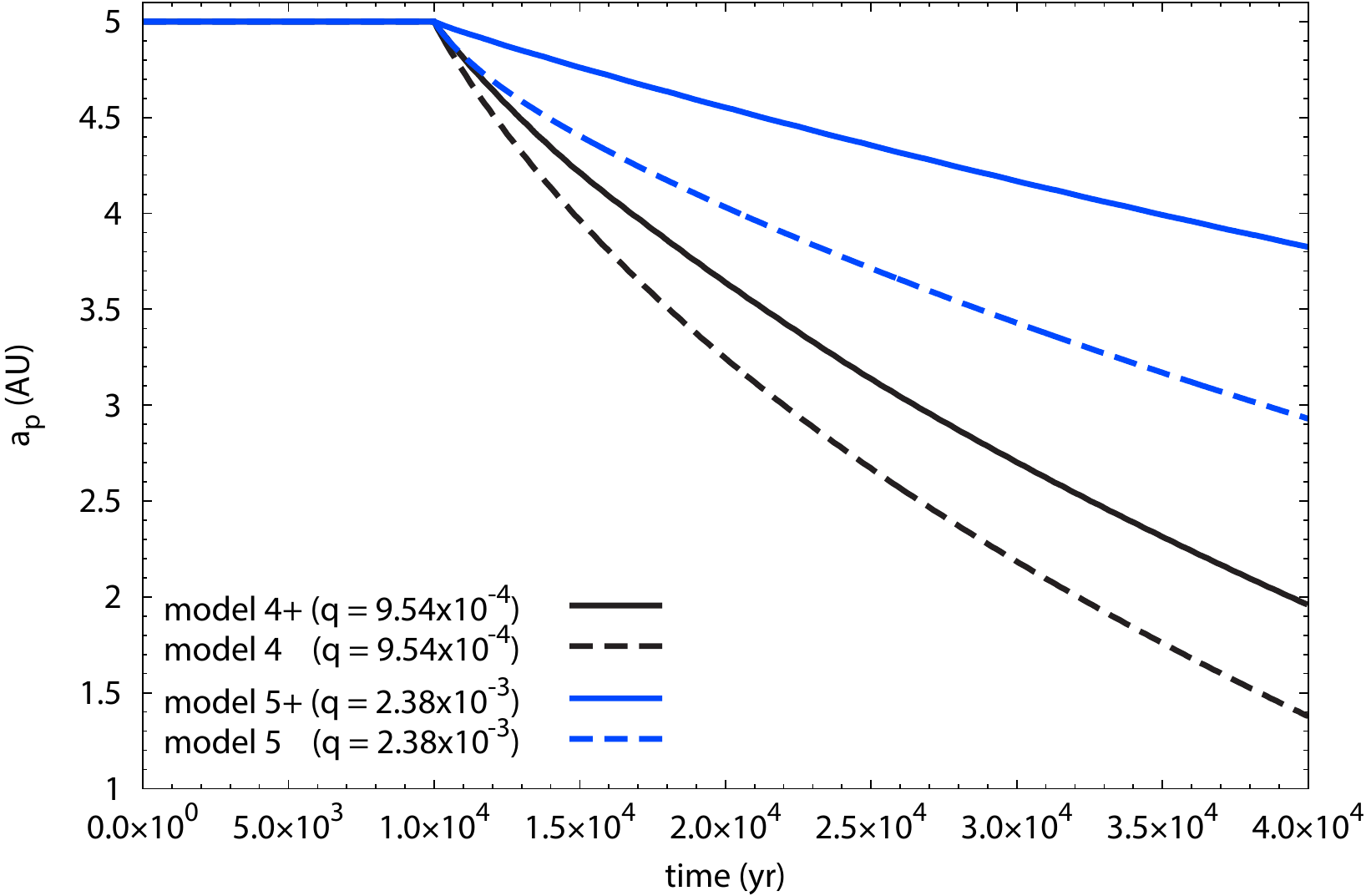}}
  \caption{Migration history of high-mass planets in models~4 and 5. Solid and dashed lines correspond to models in which the indirect potential is taken into account and neglected, respectively.}
  \label{fig:model-5-6}
\end{figure}

The effect of stellar motion on the radial migration of high-mass planets is investigated in models~4 and 5 assuming $\alpha=5\times10^{-3}$. In these models, the embedded planets  with $q=9.54\times10^{-4}$ and $2.38\times10^{-3}$ (corresponding to $M_\mathrm{p}=1\,M_\mathrm{Jup}$ and $2.5\,M_\mathrm{Jup}$, respectively) are initially on circular orbits. According to our simulations, no significant planetary and disk (local or global) eccentricities develop for these planetary masses throughout the simulations. The planets are allowed to migrate only after the formation of a fully fledged gap (after about 100 orbits) in order to model type\,II migration exclusively. Since the planets are not allowed to accrete gas, their mass remains constant throughout the simulation. The migration history is shown in Figure\,\ref{fig:model-5-6} for models~4+/4, and 5+/5.

After opening the gap, planets start migrating inwards in all models. The ratio of the local disk mass (which is the mass of the disk inside the planetary orbit) to the planetary mass is $M_\mathrm{disk}(R_0)/M_\mathrm{p}=4\pi \Sigma(R_0)R_0^2/q$, where $R_0=a_\mathrm{p}+2.5R_H(a_\mathrm{p})$.  Since $M_\mathrm{disk}(R_0)/M_\mathrm{p}>1$ as long as $a_\mathrm{p}>1.5$, the planets should migrate in the disk-dominated type II regime according to \citet{CridaMorbidelli2007}. Assuming that there is now gas flow across the planetary orbit, the change in the planetary semi-major axis as a function of time can be expressed as
\begin{equation}
        \frac{da_\mathrm{p}}{dt}=-a_\mathrm{p}\left[\frac{3}{2}\frac{\nu(R_0) }{R_0^2}\right],
        \label{eq:type-II}
\end{equation}
where $\nu(R)=\alpha c_\mathrm{s}(R)^2/\Omega(R)$.  Therefore, the migration rate is independent of the planet mass and the planet migrates on the viscous timescale. However, this is not the case 
and Figure~\ref{fig:model-5-6} demonstrates a clear dependence of the migration rate on the planetary
mass.  \citet{Duffelletal2014} have shown that the gas can cross the planetary orbit, for which case the above simple description of type\,II migration is not valid. In agreement with \citet{DurmannKley2015}, we also found that the gap-opening planet does not migrate at the viscous timescale and its migration rate is inversely proportional to its mass.

Moreover, a significant effect of stellar motion on planetary migration can be observed. If the indirect potential is included, the migration rate decreases significantly. However, as the planet's orbital distance decreases the migration rate becomes less dependent on whether stellar motion is
taken into account or not.

\subsubsection{On migration torques}

Let us analyze the torques exerted on the migrating planet. The change in the semi-major axis $a_\mathrm{p}$ of the planet having a mass $q$ due to a small change in its angular momentum can be expressed as 
\citep[e.g.,][]{Vor2013}
\begin{equation}
\frac{da_\mathrm{p}}{dt}=\frac{2\Gamma}{q a_\mathrm{p} \Omega(a_\mathrm{p})},
\label{eq:consangmom}
\end{equation}
where $\Omega(a_\mathrm{p})=a_\mathrm{p}^{-3/2}$ is the angular velocity of the planet and $\Gamma$ is the torque exerted on the planet.  The torque $\Gamma$ has two sources: one coming from the 
non-axisymmetric disk and the other from the star itself if stellar motion is allowed. 

The disk torque can be calculated numerically by summing the torques exerted on the planet by the individual grid cells, that is, 
\begin{equation}
        \Gamma_\mathrm{disk}=\sum_{i,j=1}^{N_R,N_\phi}\left[Fy_{i,j}(x_\mathrm{p}-x_\mathrm{bc})-Fx_{i,j}(y_\mathrm{p}-y_\mathrm{bc})\right],
        \label{eq:torque_disc}
\end{equation}
where $Fy_{i,j}$ and $Fx_{i,j}$ are the $x$- and $y$-components of the gravity force felt by the planet from the grid cells with coordinates $x_{i,j}$ and $y_{i,j}$. Here, $x_\mathrm{p}$, $y_\mathrm{p}$ and $x_\mathrm{bc}$, $y_\mathrm{bc}$ are the coordinates of the planet and the barycenter of the star-planet-disk system, respectively.  The components of the gravitational force of the disk felt by the planet can
be written as:
\begin{equation}
Fx_{i,j}=F_{i,j} \frac{x_{i,j}-x_\mathrm{p}}{d_{i,j}+\epsilon H(d_\mathrm{p})},
\end{equation}
and
\begin{equation}
Fy_{i,j}=F_{i,j} \frac{y_{i,j}-y_\mathrm{p}}{d_{i,j}+\epsilon H(d_\mathrm{p})},
\end{equation}
where $d_{i,j}=[(x_{i,j}-x_\mathrm{p})^2+(x_{i,j}-x_\mathrm{p})^2]^{1/2}$ is the distance of the center of grid cell $(i,j)$ to the planet. The term $\epsilon H(d_\mathrm{p})$ appears due to the applied smoothing of the planetary potential. Taking into account that we also use a torque cut-off (see details  in Section\,\ref{sec_GFARGO}), the force exerted by the individual disk cells, $F_{i,j}$, can be given as
\begin{equation}
F_{i,j}=\frac{A_{i,j}\Sigma_{i,j}\left(1-\exp{(-(d_{i,j}/R_\mathrm{H})^2)} \right)}{\left(d_{i,j}+\epsilon H(d_\mathrm{p})\right)^2},
\end{equation}
where $A_{i,j}$ and $\Sigma_{i,j}$ are the surface area and surface mass density of the grid cell $(i,j)$, respectively. 
 
If the barycenter of the star-disk-planet system is not centered on the star, a non-zero torque is exerted
on the planet by the star itself (see, e.g., \citealp{Ataieeetal2013,Regalyetal2013}). The stellar torque, $\Gamma_\mathrm{*}$, can be given as 
\begin{equation}
        \Gamma_*=Fy_*(x_\mathrm{p}-x_\mathrm{bc})-Fx_*(y_\mathrm{p}-y_\mathrm{bc}),
        \label{eq:stellar_torque1}
\end{equation}
where
\begin{equation}
        Fx_*=\frac{q}{(d_\mathrm{p}+\epsilon H(d_\mathrm{p})^2}\frac{-x_\mathrm{p}}{d_\mathrm{p}+\epsilon H(d_
\mathrm{p})}        
        \label{eq:stellar_torque2}
\end{equation}
and 
\begin{equation}
        Fy_*=\frac{q}{(d_\mathrm{p}+\epsilon H(d_\mathrm{p})^2}\frac{-y_\mathrm{p}}{d_\mathrm{p}+\epsilon H(d_\mathrm{p})}.
        \label{eq:stellar_torque3}
\end{equation}
Since we use a frame of reference that co-rotates with the planet, 
$y_\mathrm{p}$ becomes equal to zero and Equation~(\ref{eq:stellar_torque3}) vanishes, resulting in 
\begin{equation}
\Gamma_*=-\frac{q}{(d_\mathrm{p}+\epsilon H(d_\mathrm{p}))^3}x_\mathrm{p}y_\mathrm{bc}.
\end{equation}

Now, let us compare the total torque $\Gamma_\mathrm{tot}=\Gamma_\mathrm{disk}+\Gamma_\mathrm{*}$ felt by the planet in models with and without stellar motion. In models without stellar motion, $x_\mathrm{bc}=0$
and $y_\mathrm{bc}=0$. In the co-rotating frame $y_\mathrm{p}=0$, thus the second term of Equation~(\ref{eq:torque_disc}) and obviously $\Gamma_*,$ vanish,  resulting in the following expression for the total torque 
\begin{equation}
\Gamma^\mathrm{NI}_\mathrm{tot}=\sum_{i,j}^{N_r,N_\phi}\frac{A_{i,j}\Sigma_{i,j}\left(1-\exp{(-(d_{i,j}/R_\mathrm{H})^2)} \right)}{\left(d_{i,j}+\epsilon H(d_\mathrm{p})\right)^3}y_{i,j}x_\mathrm{p},
\end{equation}
where the index "NI" means that the indirect potential is not taken into account.
However, if stellar motion is taken into account, the second term in Equation~(\ref{eq:torque_disc}) and the stellar torques are non-zero. In this case, the total torque can be given as
\begin{equation}
\Gamma_\mathrm{tot}=\Gamma^\mathrm{NI}_\mathrm{tot}\left(1-\frac{x_\mathrm{bc}}{x_\mathrm{p}}\right)+\sum_{i,j}^{N_r,N_\phi}\left(\frac{x_{i,j}-x_\mathrm{p}}{y_{i,j}}\right)y_\mathrm{bc}+\Gamma_*.
\label{eq:gamma_tot}
\end{equation}
We emphasize that here we implicitly assume that the gas density distribution is independent of whether the stellar motion is on or off. With a plausible assumption of $x_\mathrm{bc}>0$ in our models,\footnote{In a co-rotating frame ($x_\mathrm{p}>0$), the barycenter of the star-planet-disk system is always located between the star and the planet, unless the barycenter shift due to the disk  is larger than the barycenter shift due to the planet, which would require a very massive disk.} the second term inside the first bracket in Equation~(\ref{eq:gamma_tot}) acts to slow down the planetary migration.  On the contrary, the second term and the stellar torque in Equation~(\ref{eq:gamma_tot}),
both being dependent on $y_\mathrm{bc}$ which can be either positive or negative, can either increase or decrease the planetary migration rate. We note that in the co-rotating frame the value of $y_\mathrm{bc}$ is determined solely by the barycenter of the disk itself.  

\begin{figure}
  \resizebox{\hsize}{!}{\includegraphics{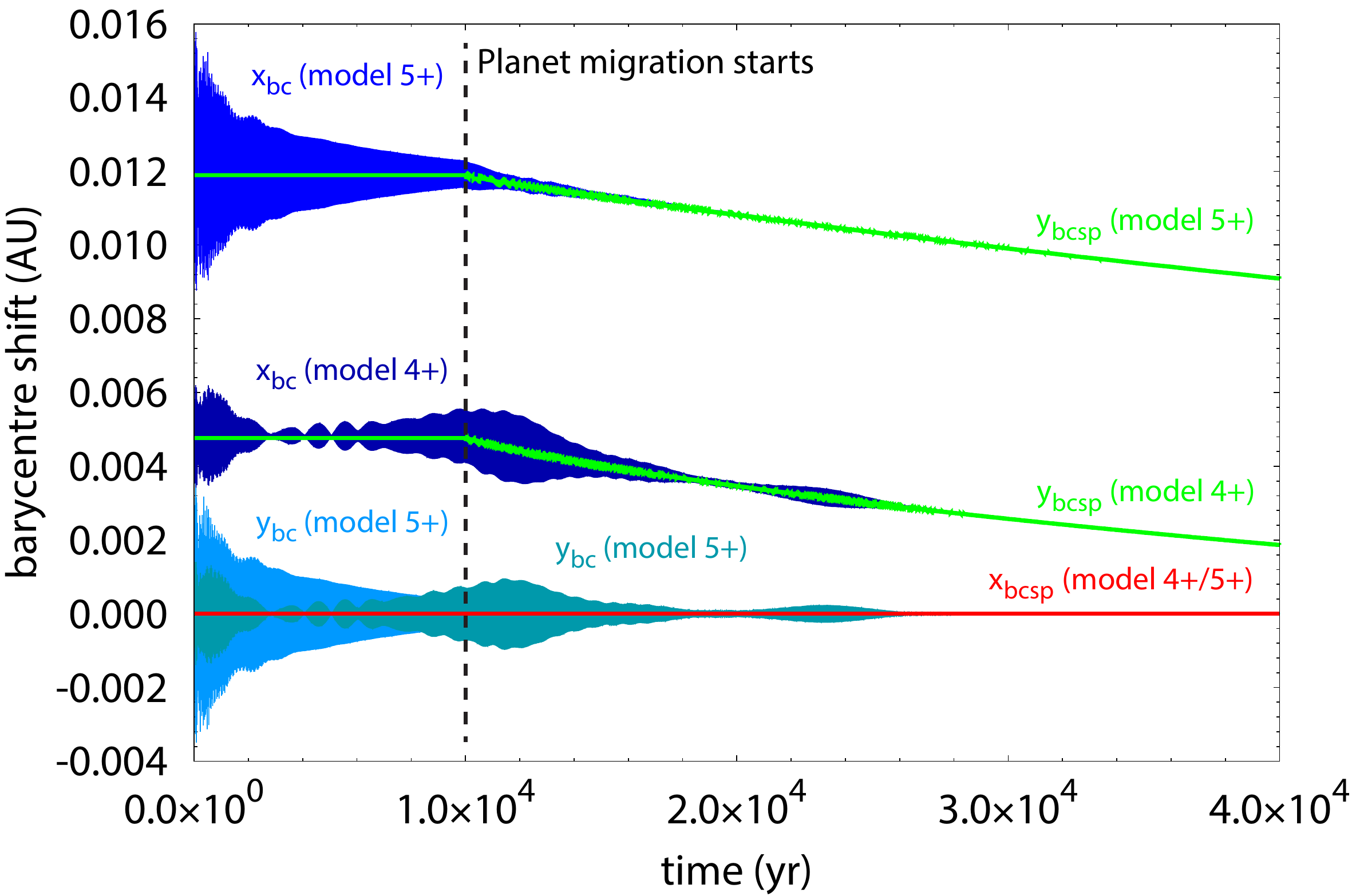}}
  \caption{Barycenter shift (in au) of the star-planet ($x_\mathrm{bc sp}$, $x_\mathrm{bc sp}$) and star-plant-disk systems ($x_\mathrm{bc}$, $x_\mathrm{bc}$) as a function of time for models 4+ and 5+. Since the shifts are highly oscillating, the running average over 50 snapshots is also shown.}
  \label{fig:model-5-6-bcshift}
\end{figure}

Figure~\ref{fig:model-5-6-bcshift} shows the barycenter coordinates of the star-planet-disk system ($x_\mathrm{bc}$, $y_\mathrm{bc}$) and the barycenter coordinates of the star-planet system ($x_\mathrm{bcsp}$, $y_\mathrm{bcsp}$) as a function of time in models 4+ and 5+.  As one can see,  both the $x_\mathrm{bc}$ and $y_\mathrm{bc}$ oscillate around the barycenter of the star-planet system. The amplitude of the oscillation decreases with time and practically vanishes by $t=10^4$\,yr (corresponding to about 1000 orbits of the planet at 5\,au). This means that the barycenter of the disk slowly converges to the barycenter of the star-planet system, even if the planet is migrating.  We emphasize that $y_\mathrm{bc}$ converges to zero, therefore both $\Gamma_*$ and the term proportional to $y_\mathrm{bc}$ in Equation~(\ref{eq:gamma_tot}) vanish. Since $x_\mathrm{bc}\simeq x_\mathrm{bcsp}=qx_\mathrm{p}/(1+q)$, the total torque becomes $\Gamma_\mathrm{tot}\simeq(1/1+q)\Gamma_\mathrm{tot}^\mathrm{NI}$, which is too small to explain the observed slowdown of migration for models 4+ and 5+. Thus, the assumption that the disk surface density distribution is the same in models 4 and 4+ and also in models 5 and 5+ must be wrong.

\begin{figure}
  \resizebox{\hsize}{!}{\includegraphics{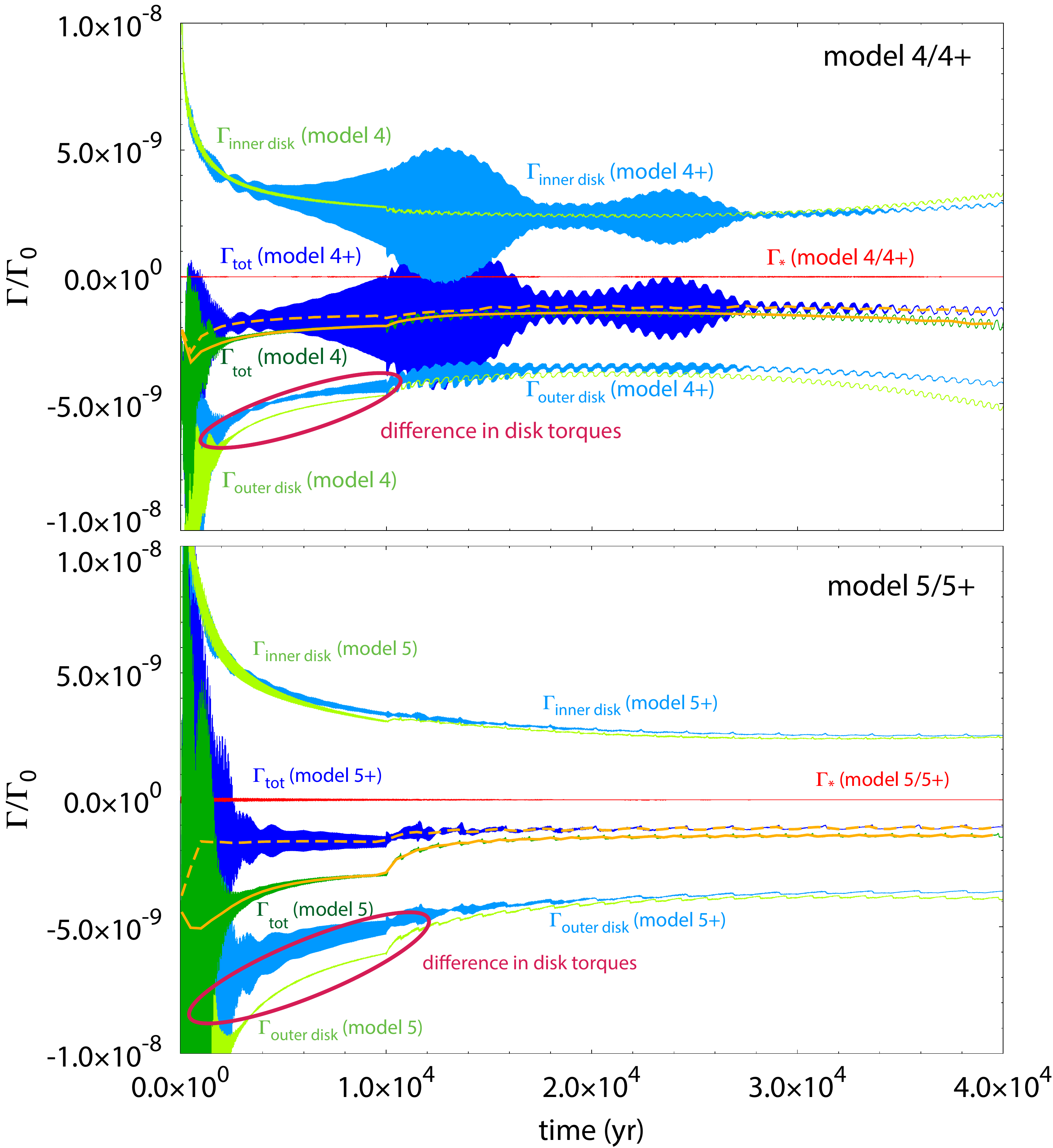}}
 \caption{Torques exerted on the planet in models 4/4+ (upper panel) and 5/5+ (lower panel). Significant departures can be observed only for the outer disk torque ($\Gamma_\mathrm{outer~disk}$), which eventually explains the decline of the total torque, $\Gamma_\mathrm{tot}$, (see orange solid and dashed curves for models 4/5 and 4+/5+, respectively) in the models where stellar motion is taken into account.}
 \label{fig:torques4-5}
\end{figure}

According to Equation~(\ref{eq:consangmom}), the change in the planet's semi-major axis is proportional to the normalized torque, $\Gamma/\Gamma_0$, where $\Gamma_0=q a_\mathrm{p}\Omega(a_\mathrm{p})$ is  the normalization factor. Figure~\ref{fig:torques4-5} shows  $\Gamma/\Gamma_0$ against time in models 4/4+ and 5/5+. It is notable that $\Gamma/\Gamma_0$ has a significantly larger magnitude in models 4
and 5 than in models 4+ and 5+ (see the solid and dashed running average curves), which explains the observed differences in the migration rates. Taking into account that $y_\mathrm{bc}\simeq0$ and 
$\Gamma_*\simeq0$, and also that the magnitude of the inner disk torque is not sensitive to the indirect potential, the difference in the migration rates can be explained by the difference in the outer disk torque alone. Since the magnitude of the difference in the outer disk torque decreases with time, the migration rates of models 4/4+ and 5/5+ slowly decrease.

\begin{figure}
  \resizebox{\hsize}{!}{\includegraphics{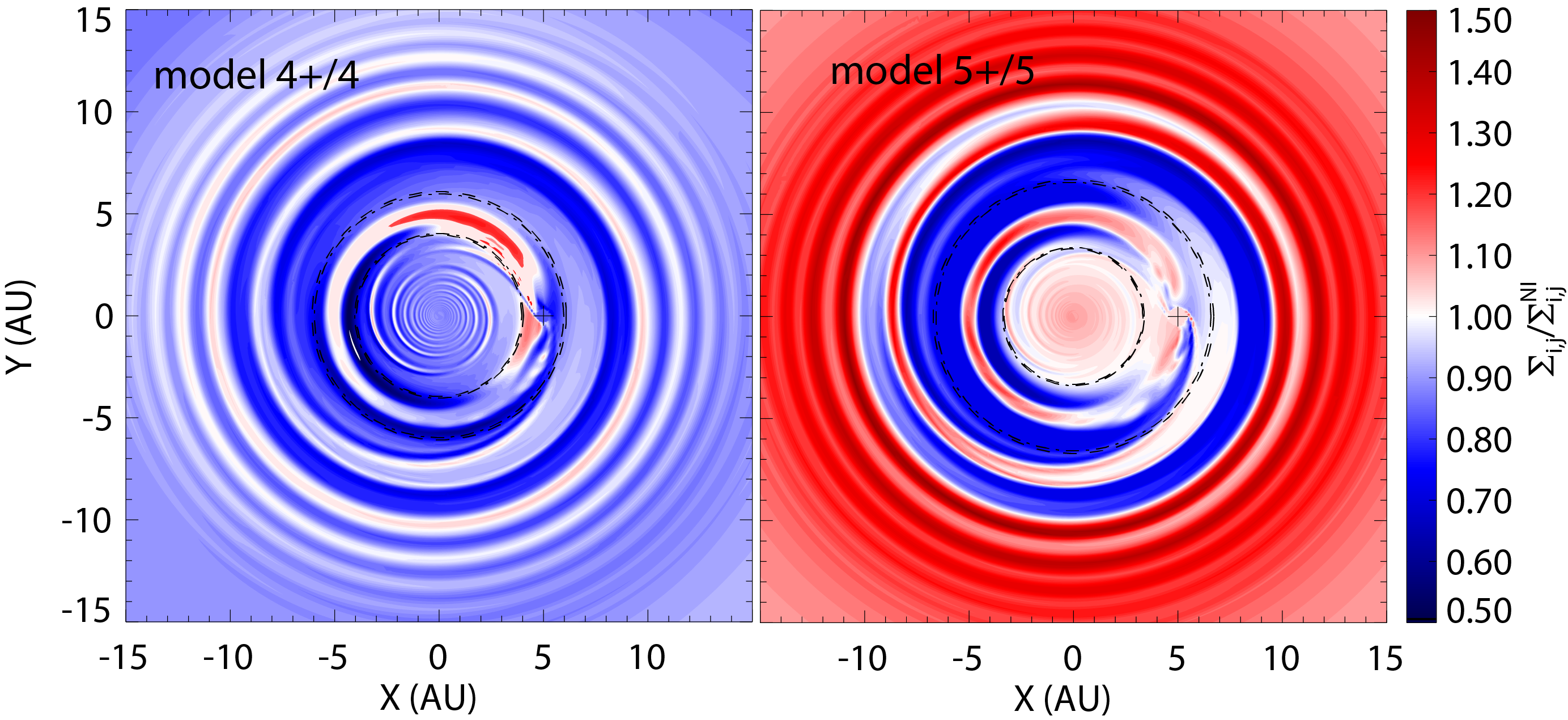}}
 \caption{Ratio of the surface mass densities in models 4+/4 (left panel), and models 5+/5 (right panel) at the beginning of the planetary migration.  The inner and outer gap edges (defined by half maximum density) are found to be nearly identical in models 4/4+ and 5/5+ shown with the dashed circles.}
 \label{fig:model5denscomp}
\end{figure}

To explain the decline of the disk torque, we plot in Figure~\ref{fig:model5denscomp} the ratio of the surface densities in model 5+ and 5 ($\Sigma_\mathrm{i,j}/\Sigma_\mathrm{i,j}^\mathrm{NI}$) at the beginning of the planetary migration. The inner disk has a slight mass enhancement ($\sim10$\%), while beyond the gap outer wall there is a significant decline ($\simeq30$\%) in density ratio in models 4+/5+ compared to models 4/5.  Although the difference in the density distributions weakens with time, it is still present at the end of the simulations. The outer disk removes angular moment from the planet, with the magnitude being proportional to the density. Therefore, the angular moment removal is smaller in model 4+ and 5+ than in model 4 and 5, respectively, which explains the observed migration rates.

In summary, it is neither the stellar torque nor the barycenter shift that causes a slowdown of
the migration rate in models with stellar motion. Actually, the planetary migration slows down because the outer disk, with respect to the planetary orbit, contains less mass, which results in 
a less efficient angular moment removal from the planet.

\subsection{Large-scale vortex formation (model~6)}
\begin{figure}
  \includegraphics[width=\columnwidth]{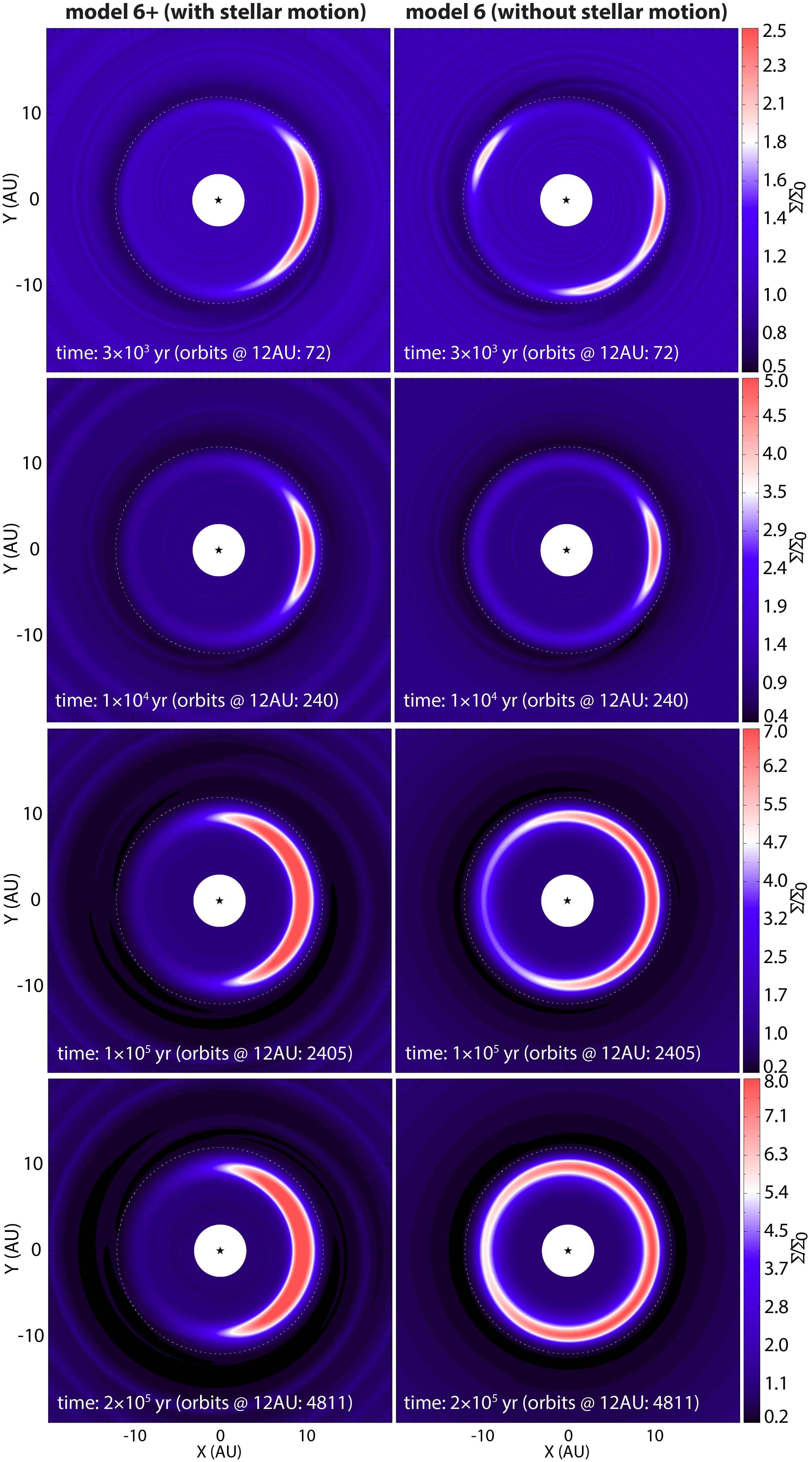}
  \caption{Evolution of large-scale vortex at viscosity transition in model~6. Left and right columns correspond to simulations in which the indirect potential is taken into account (model~6+) and neglected (model~6), respectively. The density distribution ($\Sigma$) is normalized by the initial one  ($\Sigma_0$).}
  \label{fig:model-7}
\end{figure}

\begin{figure}
  \resizebox{\hsize}{!}{\includegraphics{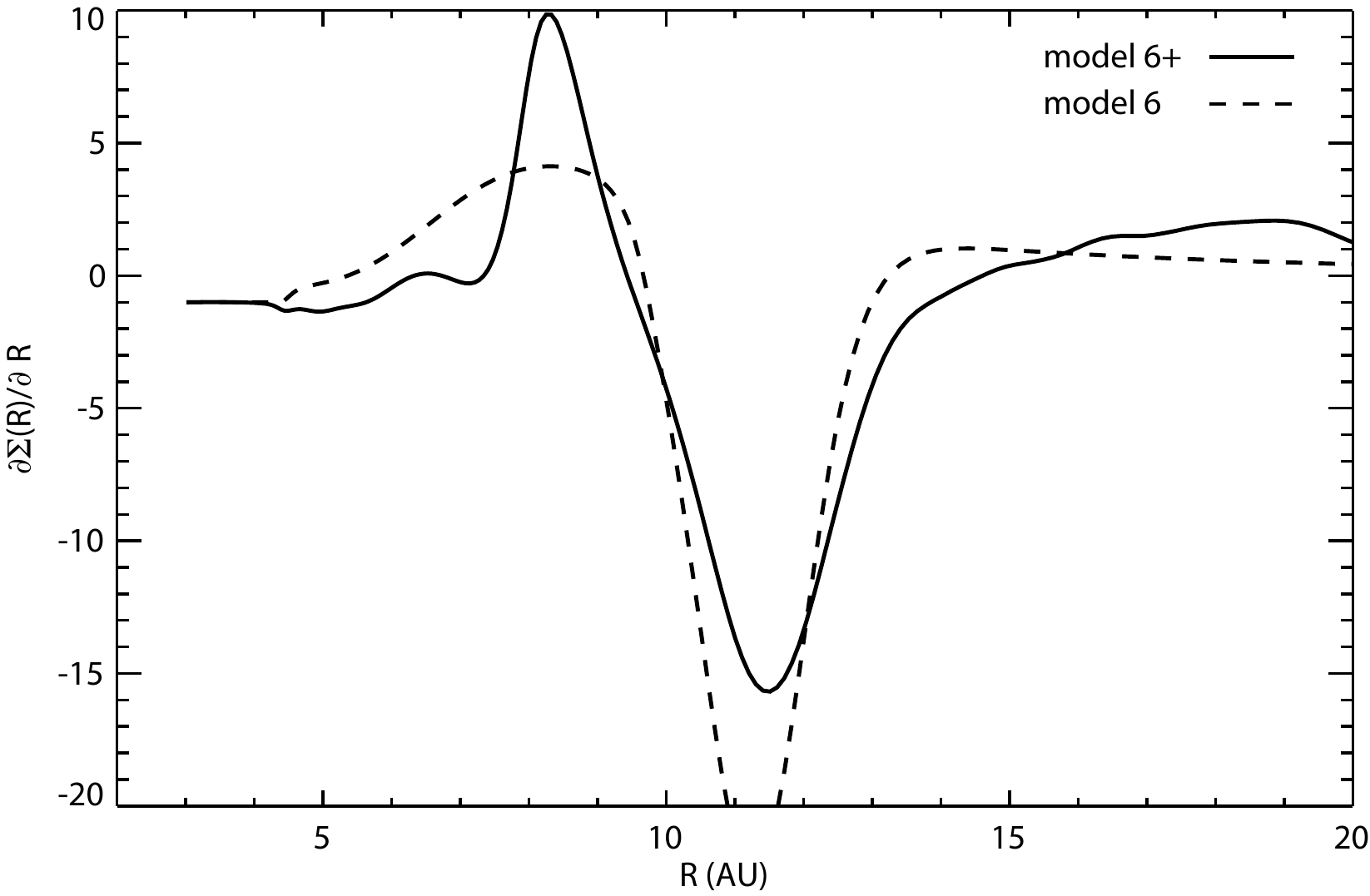}}
  \caption{Density gradient profile at the end of simulations for models~6+ and 6.}
  \label{fig:model-7-gradient}
\end{figure}

The formation and evolution of a large-scale vortex is investigated in model~6, in which a sharp viscosity transition is set by Equation\,(\ref{eq:deltaalpha}) at $R_\mathrm{dze}=12$\,au. Due to the sudden change in the accretion rate at the viscosity transition, a sufficiently sharp density enhancement forms, where the disk becomes unstable to the RWI \citep{VarniereTagger2006,Terquem2008}. The fastest growing mode is found to be $m = 5$ in our models, thus initially five anticyclonic vortices form which later coalesce \citep{Lovelaceetal1999,Lietal2000}. As a result, a single, large-scale vortex develops at the viscosity transition.

Because the density bump formed at the viscosity transition strengthens with time, the RWI excitation criterium remains fulfilled until the end of simulation. As a result, the vortex is long-lived (may be present for the disk lifetime) as was shown in \citet{Regalyetal2012} and later confirmed by \citet{Lin2014}.

Figure\,\ref{fig:model-7} shows four snapshots of the disk inner region (density distribution normalized by the initial one) for model~6+ (left column) and model~6 (right column). The excitation of the RWI can be observed in both models by about $1.5\times10^3$\,yrs. The vortex coagulation, that is, the formation of the $m=1$ large-scale vortex (upper row of Figure\,\ref{fig:model-7}) is faster in model~6+ than in model~6. However, a fully fledged single vortex develops in both simulations by $\sim10\times10^3$\,yrs (corresponding to about 75 orbits at 12\,au). Generally, the large-scale vortex is stronger in model~6+ than in model~6 throughout the simulation. By the end of the simulation, the large-scale vortex practically transformed to a ring-like density structure in model~6. In contrast, the vortex lasts to the end of the simulation in model~6+.

Figure\,\ref{fig:model-7-gradient} shows the gradient of the azimuthally averaged radial density profile ($\partial\Sigma(R)/\partial R$) at the end of simulation for models~6+ and 6. It is evident that the density bump is not sharp enough at $R\simeq10$\,au (at a radial distance of the vortex center) to maintain the vortex for model~6, which eventually causes the slow dissolution of the vortex.

\begin{figure}
  \resizebox{\hsize}{!}{\includegraphics{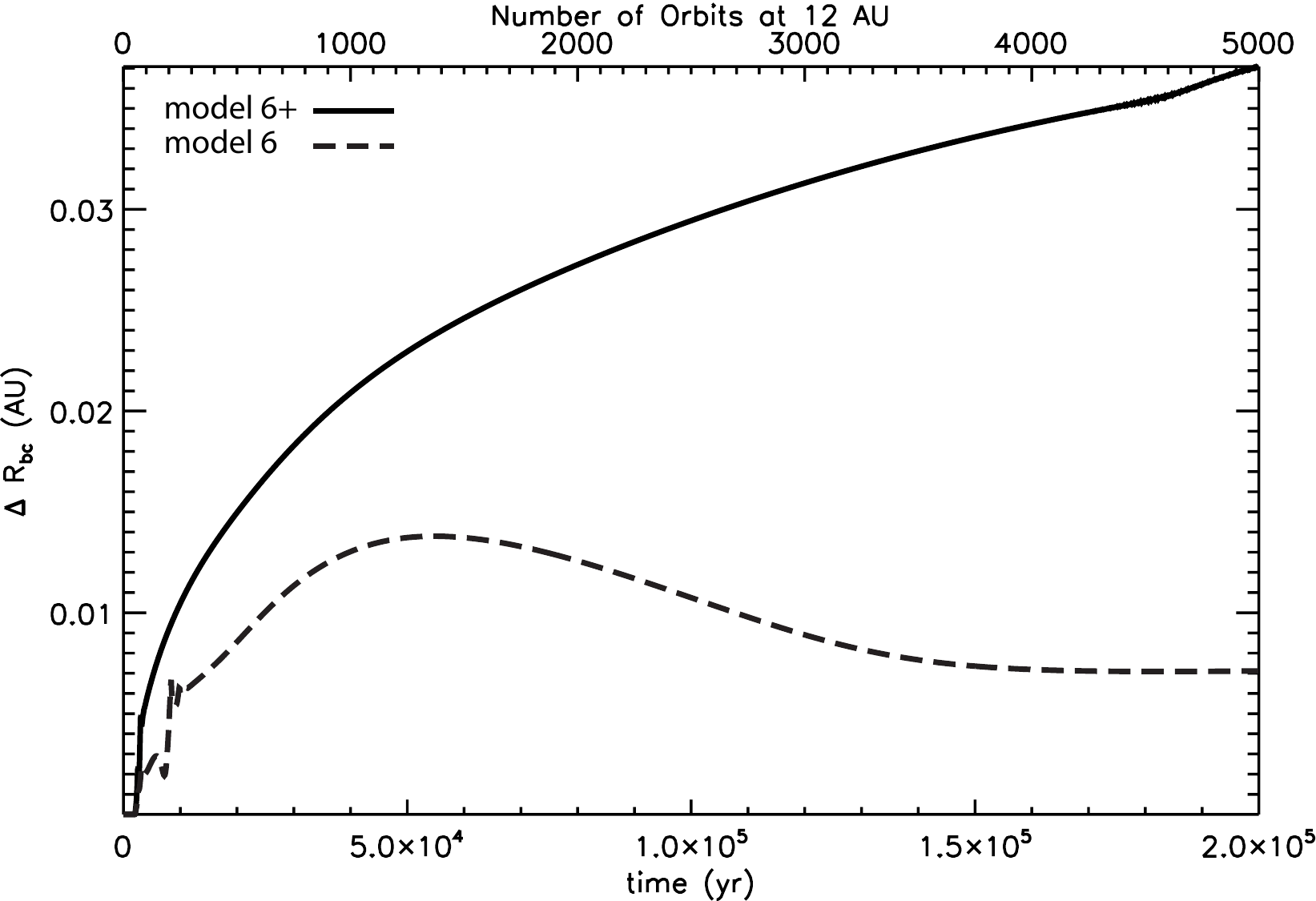}}
  \caption{Barycenter shift against time in models~6+ and 6. Solid and dashed lines correspond to models in which the indirect potential is taken into account and neglected, respectively.}
  \label{fig:model-7-bcshift}
\end{figure}

Figure\,\ref{fig:model-7-bcshift} shows the distance between the barycenter of the system and the star as a function of time. Evidently, the barycenter shift  is significant in model~6+, whereas it is much smaller in model~6, reaching its maximum ($\sim0.014$\,au) at $5\times10^4$\,yr. It is also evident that the magnitude of the barycenter shift slowly decreases after reaching its maximum in model~6, whereas it continuously increases in model~6+. We note, however, that the effect of the barycenter shift in model~6 has no effect on the evolution of the vortex; it only reflects the asymmetry in the density distribution. 

We note that \citet{MittalChiang2015} proposed that a large-scale vortex can be sustained in massive disks due to an artificial  displacement of the star from the barycenter of the system. However, they applied an indirect potential artificially. Recently, \citet{ZhuBaruteau2016} found that the indirect potential indeed promotes the formation of a large-scale vortex, however, in their simulation an artificially imposed axisymmetric Gaussian density bump served as the background of the density gradient that excites the RWI. Here we not only confirm the above findings but also observe that the stellar motion promotes the excitation of the RWI.

\section{Conclusions}

In this paper, we presented a detailed investigation of the effect of stellar motion on the evolution of protostellar and protoplanetary disks by means of 2D numerical hydrodynamic simulations. The stellar motion is caused by various non-axisymmetric structures in the disk and infalling envelope, which emerge during the early and late evolutionary phases. Two distinct codes were used: FEoSaD for protostellar and GFARGO for protoplanetary disks. Both codes employ the grid-based cylindrical coordinate system centered on the star. The motion of the central star is taken into account by calculating the indirect potential, which occurs when the equations of motion for the disk are formulated in the non-inertial frame of reference moving with the star.

In the protostellar disks, asymmetries can develop due to disk gravitational 
fragmentation resulting in the formation of massive clumps and spiral arms. In the 
protoplanetary disks, asymmetries can develop due to an embedded planet or large-scale vortex forming, for example, at viscosity transitions. In both cases, significant effects of the stellar motion were found. Our key results for the early disk evolution are as follows:
\begin{itemize}
\item[1)]  When the stellar motion is taken into account, part of the envelope angular 
momentum is transferred to the orbital motion 
of the protostar around the center of mass, reducing the net angular momentum of the forming 
disk. This leads to a somewhat smaller disk mass and radius, and, 
as a consequence, to a weaker gravitational instability and fragmentation as compared to the 
case without stellar motion.
\item[2)] The character of mass accretion rate is also moderately affected by the stellar motion, leading
to a mild reduction in the accretion variability inherent to gravitationally unstable disks.
This is consistent with the overall reduction in the disk mass and size, and in 
the strength of the gravitational instability of the disk. A Fourier analysis 
of the spiral structure shows that models with stellar motion are dominated by one-armed 
spiral modes, while other low-order modes have roughly a similar strength.
\item[3)]  Stellar motion has a profound effect on the collapsing envelope, changing its shape 
from an initially axisymmteric state to a profoundly non-axisymmetric configuration.
This causes a significant excursion of the center of mass of the system,
complemented with small-scale wobbling due to the gravitational perturbation of the disk.
\end{itemize}
\noindent
Our key results for the protoplanetary disk evolution are as follows:

\begin{itemize}
\item[4)] The gap formation is significantly slower if the stellar motion is neglected: after about 100 orbits of  a non-migrating $5\,M_\mathrm{Jup}$ planet, the gap is only partially depleted (the planetary horseshoe region is filled with gas). On the contrary, the gap is fully depleted if the stellar motion is taken into account. However, the gap is completely devoid of gas after about several hundred orbits independent of whether the stellar motion is on or off. The time required for the horseshoe gas to 
vanish is inversely proportional to the disk viscosity.
\item[5)] Independent of the applied viscosity, the gap carved by a non-migrating $5\,M_\mathrm{Jup}$
planet is significantly narrower, if the stellar motion is taken into account. If the density gradient at the outer gap edge is large enough for a long time (which requires low viscosity, e.g., $\alpha\leq5\times10^{-4}$) a long-term large-scale vortex forms. However,  we observed the vortex formation only if the stellar motion is taken into account. 
\item[6)] The azimuthally averaged disk eccentricity (excited by a non-migrating $5\,M_\mathrm{Jup}$ planet) is only slightly modified by the effect of stellar motion. However, the time required for the formation of a quasi-steady eccentric disk is shorter in models with stellar motion.
\item[7)] The radial migration of high-mass planets (investigated in the range of $1.0-2.5\,M_\mathrm{Jup}$) is also affected by the stellar motion: the type\,II migration rate is significantly smaller if the effect of stellar motion is taken into account. This phenomenon can be explained by the fact that the gap outer edge is depleted in gas if the stellar motion is taken into account, resulting in 
a smaller magnitude of the outer disk torque.
\item[8)] Regarding the large-scale vortex formation at a sharp viscosity transition, we found that the formation of the RWI unstable density bump and the vortex coagulation are stimulated by the stellar motion. If the stellar motion is included, the fully fledged large-scale vortex is stronger (the azimuthal asymmetry is larger). On the contrary, the large-scale vortex dissolves (only a ring-like perturbation remains) within about 5000 orbits at the distance of the vortex center in models where the effect of the stellar motion is neglected.
\end{itemize}

Our general conclusion is that the stellar motion has a notable effect on the evolution of 
gravitationally unstable protostellar disks and a strong effect on the planet- or large-scale 
vortex-bearing protoplanetary disks. Therefore, inclusion of the indirect potential 
is recommended in grid-based hydrodynamics simulations of circumstellar disks.

Finally, we note that special care with the computational boundaries needs to be taken
when implementing the indirect potential, especially when the mass of the circumstellar disk is commensurable to that of the central star. The acceleration of the non-inertial frame of 
reference may lead to an artificial distortion of the disk or/and the infalling envelope, 
especially near the outer computational boundary if the numerical resolution is insufficient, 
as is the case for logarithmically spaced grids. To mitigate this problem,
we extended the computational domain and filled it with a low density external environment.
However, a better solution needs to be found in the future, probably involving the introduction
of moving grids or boundaries.

\section{Acknowledgements}
The authors are thankful to the referee for insightful comments.
The project was partly supported by the joint OeAD-OMAA program through project 90\"ou25 and by the Hungarian OTKA Grant No. 119993. Zs. R. acknowledges support from Momentum grant of the MTA CSFK Lend\"ulet Disk Research Group K101393 and NVIDIA Corporation for donating of the Tesla 2075 and K40 GPUs used for this research.
E. I. Vorobyov acknowledges support from the Austrian Science Fund (FWF) under research grant I2549-N27. The simulations were performed on the Vienna
Scientific Cluster (VSC-2 and VSC-3) and on the
Shared Hierarchical Academic Research Computing Network (SHARCNET).

\end{document}